\newif{\ifarxiv}
\newif{\ifdraft}
\newif{\ifremarks}

\arxivtrue  
\drafttrue  
\remarkstrue  

\ifremarks
\newcommand{\remarkcb}[1]{{\renewcommand{\bfdefault}{b}\color{cyan}{\textbf{C:~#1}}}}
\newcommand{\remarktb}[1]{{\renewcommand{\bfdefault}{b}\color[RGB]{0,150,0}{\textbf{T:~#1}}}}
\newcommand{\comG}[1]{{\renewcommand{\bfdefault}{b}\color{olive}{\textbf{G:~#1}}}}
\newcommand{\comP}[1]{{\renewcommand{\bfdefault}{b}\color{magenta}{\textbf{P:~#1}}}}
\fi
\providecommand{\remarkcb}[1]{\ignorespaces}
\providecommand{\remarktb}[1]{\ignorespaces}
\providecommand{\comG}[1]{\ignorespaces}
\providecommand{\comP}[1]{\ignorespaces}

\documentclass[
 reprint,
 amsmath,
 amssymb,
 aps,
 prl,
 nofootinbib,
 nobibnotes,
 longbibliography,
 superscriptaddress,
 preprintnumbers
]{revtex4-2}

\ifdraft
\fi

\ifarxiv


\makeatletter
\def\paragraph{%
  \@startsection
    {paragraph}%
    {4}%
    {\z@}%
    {1.5ex \@plus.5ex \@minus .2ex}%
    {-1em}%
    {\normalfont\normalsize\bfseries}%
}%
\makeatother

\fi

\usepackage[a4paper,vmargin=2cm,hmargin=1.5cm]{geometry}
\usepackage{amsmath}

\usepackage[utf8]{inputenc}
\usepackage{comment}
\usepackage[pdftex]{graphicx}
\usepackage{wasysym}
\usepackage{amsfonts}
\usepackage{ifsym}

\usepackage[dvipsnames]{xcolor}
\usepackage{color}
\usepackage{dcolumn}
\usepackage{bm}
\usepackage{hyperref}

\usepackage{graphbox}
\usepackage{tikz}

\hypersetup{plainpages=false}
\hypersetup{pdfpagemode=UseOutlines}
\hypersetup{bookmarksnumbered=true}
\hypersetup{bookmarksopen=true}
\hypersetup{pdfstartview=FitH}
\hypersetup{colorlinks=true}
\hypersetup{citecolor=[rgb]{0 .4 0}}
\hypersetup{urlcolor=[rgb]{.4 0 0}}
\hypersetup{linkcolor=[rgb]{0 0 .5}}

\newcommand{\namedref}[2]{\hyperref[#2]{#1~\ref*{#2}}}
\newcommand{\secref}[1]{\namedref{Section}{#1}}
\newcommand{\appref}[1]{\namedref{Appendix}{#1}}

\newcommand{\figref}[1]{\namedref{Figure}{#1}}

\allowdisplaybreaks
\usepackage{enumitem}

\def\etal.{et\penalty50\ al.}
\usepackage{xspace}
\newcommand*{\eg}{e.\,g.\@\xspace}
\newcommand*{\ie}{i.\,e.\@\xspace}

\makeatletter\newcommand*{\etc}{%
    \@ifnextchar{.}%
        {etc}%
        {etc.\@\xspace}%
}\makeatother

\def\clap#1{\hbox to 0pt{\hss#1\hss}}

\usepackage[mathbold,autobold,greekcaps,greeklower]{mathfixs}

\usepackage{empheq}
\newlength{\widefboxpadding}
\newcommand{\nn}{\nonumber}

\newcommand{\alg}[1]{\mathfrak{#1}}
\newcommand{\grp}[1]{\mathrm{#1}}

\newcommand{\alPSU}{\alg{psu}}

\RequirePackage[extdef]{delimset}
\makeatletter
\providecommand{\brkleft}[1][r]{\begingroup\def\dlm@use{\delim(.}%
\if r#1 \def\dlm@use{\delim(.}\fi%
\if s#1 \def\dlm@use{\delim[.}\fi%
\if c#1 \def\dlm@use{\delim\{.}\fi%
\if a#1 \def\dlm@use{\delim<.}\fi%
\expandafter\endgroup\dlm@use}
\providecommand{\brkright}[1][r]{\begingroup\def\dlm@use{\delim.)}%
\if r#1 \def\dlm@use{\delim.)}\fi%
\if s#1 \def\dlm@use{\delim.]}\fi%
\if c#1 \def\dlm@use{\delim.\}}\fi%
\if a#1 \def\dlm@use{\delim.>}\fi%
\expandafter\endgroup\dlm@use}
\makeatother

\newcommand{\ee}{\mathrm{e}}
\newcommand{\ii}{i}

\newcommand{\svx}{\mathsf{x}}
\newcommand{\svy}{\mathsf{y}}
\newcommand{\lM}{\mathcal{M}}

\newcommand{\superN}{\mathcal{N}}

\newcommand{\Integers}{\mathbb{Z}}

\newcommand{\Complex}{\mathbb{C}}

\newcommand{\dd}[2][]{\mathinner{\mathrm{d}\ifx#1\empty\else{^#1}\fi#2}}
\DeclareMathDelimiterSet{\db}[1]{%
  \selectdeliml{(}\kerndelim{-4}\selectdelim[o]{(}{#1}%
  \selectdelim[o]{)}\kerndelim{-4}\selectdelimr{)}%
}
\DeclareMathDelimiterSet{\dab}[1]{%
  \selectdeliml{\langle}\kerndelim{-4}\selectdelim[o]{\langle}{#1}%
  \selectdelim[o]{\rangle}\kerndelim{-4}\selectdelimr{\rangle}%
}

\begin{document}

\preprint{DESY-26-045}

\title{Structure Constants from Q-Systems and Separation of Variables}

\author{Till Bargheer}
\affiliation{Deutsches Elektronen-Synchrotron DESY, Notkestr.~85, 22607 Hamburg, Germany}

\author{Carlos Bercini}
\affiliation{Department of Mathematics, King's College London, The Strand, London WC2R 2LS, UK}

\author{Gabriel Lefundes}
\affiliation{Universit\'e Paris--Saclay, CNRS, CEA, Institut de Physique Th\'eorique, 91191 Gif-sur-Yvette, France}
\affiliation{ICTP South American Institute for Fundamental Research, IFT-UNESP, S\~ao Paulo, SP 01440-070, Brazil}

\author{Paul Ryan}
\affiliation{Deutsches Elektronen-Synchrotron DESY, Notkestr.~85, 22607 Hamburg, Germany}

\begin{abstract}

We introduce a novel method to compute structure constants from Q-functions in
the scalar sector of planar \mbox{$\superN=4$} super Yang--Mills (SYM) and related theories.
The method derives from operatorial as well as functional separation
of variables, and the structure constants are expressed as
determinants of matrices whose entries are integrals over products of
Q-functions.
In this framework, each operator is twisted by an external angle,
mirroring the cusped Maldacena--Wilson loop.
The structure constants of local single-trace operators
in $\superN=4$ SYM are recovered in the
untwisting limit, where we obtain a one-to-one correspondence between
our key building blocks and those of the Hexagon formalism.
Retaining appropriate twists, our structure constants also perfectly
match those of the orbifold points of $\superN=4$ SYM.
Our results thus far are valid at leading order in the weak-coupling
expansion, but their formulation in terms of Q-functions provides a
natural starting point for including loop corrections.
Many of the methods we develop in this work apply more
generally to the computation of correlation functions in integrable models.
\end{abstract}

\maketitle

\section{Introduction}

Solving the dynamics of an interacting 4d QFT analytically remains an
outstanding challenge in theoretical physics. In the most symmetric
setting of $\superN=4$ super Yang--Mills theory (SYM), integrability
has come a long way towards this goal. Most notably, the Quantum
Spectral Curve (QSC)~\cite{Gromov:2013pga,Gromov:2014caa} efficiently
captures the spectra of local \cite{Gromov:2023hzc} and non-local operators
\cite{Ekhammar:2025vig} at finite coupling. On the contrary, dynamical
quantities are far less understood thus far. Extending the
integrability approach to higher-point correlators remains a central
open problem. The state of the art for three-point functions is an
expansion in string worldsheet excitations via hexagon form
factors~\cite{Basso:2015zoa}, which is very powerful for large-charge
operators, but requires re-summations that cannot be
performed in practice for small charges~\cite{Basso:2017muf}.

An alternative formalism, where
correlators are integrals over the Q-functions naturally associated to the
external operators by the QSC, would be most desirable. A natural
approach is the Separation of Variables (SoV)
framework~\cite{Sklyanin:1984sb}, where all
wave functions factorize into single-particle Q-functions, and
correlators simplify drastically~\cite{Giombi:2018qox,Cavaglia:2018lxi}.
The SoV approach to structure constants thus far has been restricted
to bosonic rank-one sectors~\cite{Jiang:2015lda,Bercini:2022jxo,Bargheer:2025kli}.
A finite-coupling formalism will necessarily have to include the full
$\alg{psu}(2,2|4)$ state space, which requires two major steps: a
generalization to higher rank, and the inclusion of fermions.

In this work, we accomplish one of these major steps by combining
operatorial~\cite{Gromov:2016itr,Maillet:2018bim,Ryan:2018fyo} and
functional~\cite{Cavaglia:2019pow,Gromov:2019wmz,Gromov:2020fwh} SoV
to derive a formalism that allows to express the leading-order
structure constant of any operator in the full $\alg{su}(4)=\mathfrak{so}(6)$ scalar
sector with any two scalar BPS operators in terms of determinants of
Q-functions.

A key aspect of our analysis are twists, which break the $\alg{su}(4)$
symmetry and thereby lift all degeneracies. Hence our result applies
uniformly to all highest-weight as well as descendant operators.
Moreover, the twists allow us to access deformations of $\superN=4$
SYM, such as orbifold points~\cite{Kachru:1998ys}.

\begin{figure}[t]
    \includegraphics[lmargin=-49mm,scale=0.75]{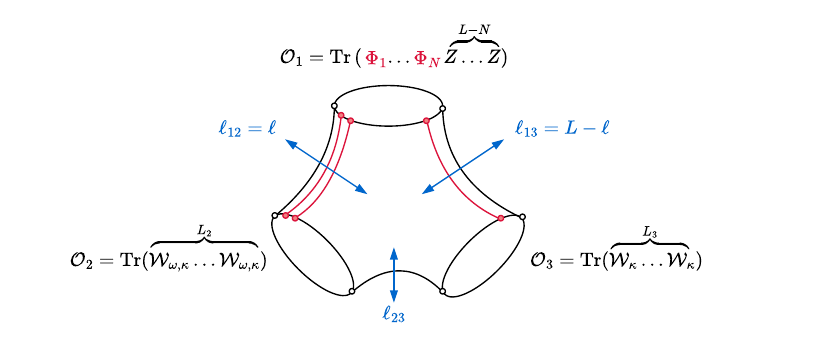}
    \vspace{-2em}
    \caption{Twisted three-point function of one excited operator of size
    $L_1$ with $N$ excitations and two protected operators of sizes $L_2$ and $L_3$. The red lines represent the propagators of the excitations $\Phi_i \in \{X,Y,\bar{X},\bar{Y},\bar{Z}\}$ coming from the non-BPS operator. The bridge lengths $\ell_{ij} = (L_i+L_j-L_k)/2$ denote the number of propagators between operators $\mathcal{O}_i$ and $\mathcal{O}_j$. In particular, we use $L_1\equiv L$,
    $\ell_{12} = \ell$ and $\ell_{13}=L-\ell$ to denote the length and bridges of the excited operator.}
    \label{fig:3Points}
\end{figure}

\section{Setup}
\label{SecSetup}

We consider the three-point function of one excited operator $\mathcal{O}_1$ with twists $z_j$, $j=1,2,3,4$, and two protected operators $\mathcal{O}_2 = {\rm Tr}(\mathcal{W}_{\omega,\kappa}^{L_2})$ and $\mathcal{O}_3 = {\rm Tr}(\mathcal{W}_{\omega}^{L_3})$ depicted in \figref{fig:3Points}. The fields $\mathcal{W}_\omega$ and $\mathcal{W}_{\omega,\kappa}$ are linear combinations of the six elementary scalars $Z,X,Y,\bar{Y},\bar{X},\bar{Z}$. In the integrable spin chain picture, the operators $\mathcal{O}_i$ are identified with eigenstates of twisted conserved charges,
\begin{equation}
    \mathcal{O}_1: |\Psi\rangle\,, \qquad
    \mathcal{O}_2: \langle \mathcal{W}_{\omega,\kappa}^{L_2}|\,, \qquad
    \mathcal{O}_3: \langle \mathcal{W}_{\omega}^{L_3}|\,,
\end{equation}
where the protected operators $\mathcal{O}_2$, $\mathcal{O}_3$ are
$\grp{SU}(4)$ rotations of the vacuum $\langle Z|$ parametrized as
\begin{equation}
    \langle \mathcal{W}_A|=\langle Z|e^{\nu_2 p^A_{13}\mathbb{E}_{13}} e^{\nu_1 p^A_{23}\mathbb{E}_{23}}e^{\nu_1 p^A_{14}\mathbb{E}_{14}}e^{\nu_2 p^A_{24}\mathbb{E}_{24}}\,.
    \label{eq:WAdef}
\end{equation}
The $\nu_j$ are simple ratios of $z_j$ -- the twists of the excited
state -- while $p_{jk}^A$ parametrize the polarizations of the
operators, both written explicitly in \appref{appReviewSU4}.

The polarizations are not completely independent due to the reduction
$p_{jk}^{\omega,\kappa=0}=p_{jk}^{\omega}$. The reference
polarizations $p_{jk}\equiv p_{jk}^{\omega=0}$ can be freely chosen up to the constraint
$p_{13}p_{24} = p_{14} p_{23}$. We emphasize that while the
definitions of these fields may seem rather ad-hoc, they arise
naturally in the SoV framework, see
\secref{sec:derivation} below.

In planar $\mathcal{N}=4$ SYM with $z_j=1$, the global $\alg{su}(4)$ R-symmetry allows
one to rotate to a canonical frame where the BPS operators are~\cite{Drukker:2009sf,Basso:2015zoa}
\begin{equation}
    \mathcal{O}_2 = \text{Tr}(Z+\bar{Z}+X-\bar{X})^{L_2},\quad \mathcal{O}_3 = \text{Tr}(\bar{Z})^{L_3}\,,
    \label{eq:BPScanonical}
\end{equation}
where $\mathcal{O}_3$ is called ``reservoir state'', since it absorbs
all excitations from the non-BPS operator $\mathcal{O}_1$.
In the presence of twists $z_j\neq1$, the $\alg{su}(4)$ symmetry is
broken, hence
the structure constants depend non-trivially on the parameters
$\omega$ and $\kappa$, while the freedom in choosing $p_{ij}$
reflects the residual $\alg{u}(1)^3$ symmetry. The untwisted planar
$\mathcal{N}=4$ super Yang--Mills setup is recovered in the $z_j\to1$
limit. Setting in addition $\omega=\kappa=1$ as well as $p_{jk}=1$, the BPS operators
$\mathcal{O}_2$, $\mathcal{O}_3$ reduce to~\eqref{eq:BPScanonical}. We refer to this as the \emph{untwisting limit}:
\begin{equation}
    z_i \to 1\,, \quad \kappa \to 1\,, \quad \omega \to 1\,.
    \label{UntwistingLimit}
\end{equation}

The central object is the normalized structure constant  $C_\ell
\equiv C^{\bullet\circ\circ}/C^{\circ\circ\circ}$ where
$C^{\circ\circ\circ}$ is the structure constant of three protected
operators of the same lengths. At leading order, $C_\ell$ is independent of the bottom bridge $\ell_{23}$, and it is equal to the spin chain overlap
\begin{equation}
    C_\ell = \frac{\langle\mathcal{W}_{\omega,\kappa}^\ell\otimes \mathcal{W}_\omega^{L-\ell} |\Psi\rangle}{||\Psi||}\,.
    \label{twistedOverlap2}
\end{equation}
The state $\ket{\Psi}$ is characterized by four Q-functions $Q_j$ and their duals $Q^j$ ($j=1,2,3,4$), defined as solutions to the fourth-order \emph{Baxter equation} $\mathcal{B} \cdot Q = 0$ and the dual equation $\mathcal{B}^\dagger \cdot Q^j = 0$. The Baxter operator $\mathcal{B}$ acts as follows:
\begin{equation}
\begin{split}
    \mathcal{B} \cdot Q  &= u^L (u-i)^L Q^{[4]} +u^L(u+i)^L Q^{[-4]} - \tau_0^+ Q\\
    & \quad -(u-i)^L\tau_+ Q^{[2]} - (u+i)^L\tau_-^{[-2]} Q^{[-2]}\,,
    \label{niceBaxter}
\end{split}
\end{equation}
where $f^{[a]}\equiv f(u+\tfrac{ai}{2})$ and $f^\pm\equiv f(u \pm \tfrac{i}{2})$ denote half-integer shifts. The functions $\tau_0(u)$ and $\tau_{\pm}(u)$ are the eigenvalues of the transfer matrices $t_0(u)$ and $t_\pm(u)$, which encode the complete set of $4L$ conserved quantities, see \appref{appTransfer}. Finally, the dual Baxter operator $\mathcal{B}^\dagger$ is obtained directly from $\mathcal{B}$ by swapping the eigenvalues $\tau_\pm\rightarrow \tau_\mp$.

The Q-functions are twisted polynomials of the form
\begin{equation}
    Q_j(u)=z_j^{-iu}q_j(u),\quad Q^j(u) = z_j^{iu}q^j(u)\,,
\end{equation}
where $z_j=e^{i\phi_j}$ are the twist parameters of the excited spin chain, and $q_j$ and $q^j$ are polynomials whose degrees $\lambda_j$ and $\lambda^{j} = L-\lambda_j$ encode the global $\alg{su}(4)$ quantum numbers of the excited state, see \appref{appSU4Baxter}. In terms of the Dynkin labels $[p,q,r]$ of the $\alg{su}(4)$ representation,
\begin{equation}
    p = \lambda_2-\lambda_1 \,,\quad
    q = \lambda_3-\lambda_2 \,,\quad
    r = \lambda_4-\lambda_3 \,,
    \label{eq:pqrlambdas}
\end{equation}
with the constraint $2L =\lambda_1+\lambda_2+\lambda_3+\lambda_4$.

For simplicity, we restrict to twists satisfying $z_1z_4=z_2z_3=1$, treating the general case in the Appendices.  For generic real angles $\phi_1$ and $\phi_2$, the spectrum of conserved charges is non-degenerate \cite{mukhin2014spaces,Chernyak:2020lgw}: each Hamiltonian eigenstate is uniquely associated with the four Q-functions $Q_j$. The situation simplifies further for our representation of $\alg{su}(4)$: knowing just two Q-functions, \eg $Q_1$ and $Q_2$, suffices to reconstruct all others \cite{Ryan:2018fyo}.

The variables $p_{jk}$ parametrize the $\alg{u}(1)^3\subset \alg{su}(4)$ Cartan symmetry,
thus $C_\ell$ depends on them rather trivially,
\begin{equation}\label{eqn:Ccovariance}
    C_\ell(p_{jk}) =
    |p_{13}|^{2(K-\bar{K})}|p_{14}|^{2\bar{K}}|p_{23}|^{2(M-K)}
    \times C_\ell^0\,,
\end{equation}
where $C_\ell^0\equiv C_\ell(p_{jk}=1)$, and
$K,M,\bar{K}$ are Cartan quantum numbers that are related to the
Dynkin labels~\eqref{eq:pqrlambdas} via
\begin{equation}\label{eqn:dynkinroots}
    [p,q,r] = [M-2K,L+K-2M+\bar{K},M-2\bar{K}]\,.
\end{equation}
Unless otherwise stated, we assume the choice $p_{ij}=1$ in what follows.

\section{Result}
\label{sec:result}

We now present our main result: The twisted structure constant~\eqref{twistedOverlap2} admits the following SoV formulation, factorized into simple universal building blocks:
\begin{equation}
    C_\ell^0 = \mathcal{N}_{\ell} \times
    \frac{\mathbb{W}_{\omega} \, \mathbb{A}_{\ell,\kappa}}{\sqrt{\mathbb{B}}}
    \label{CSoV}
\end{equation}
where $\mathcal{N}_\ell$ is a normalization defined below,
and the remaining factors are SoV inner products built from the
Q-functions of the excited state:
\begin{align}
    \mathbb{B} & =
    \mspace{45mu} \mathclap{\det_{(\alpha,j,k),(\beta,a,b)}} \mspace{48mu}
    \langle \mathcal{Q}_{j}^{[a]} u^{\beta-1} \mathcal{Q}^{k[b]}  \rangle_{L,\alpha}
    \label{BuSoV}
    \,,\\
    \mathbb{W}_{\omega} & =
    \mspace{45mu} \mathclap{\det_{\alpha,\beta}} \mspace{48mu}
    \langle \omega^{-iu}u^{\beta-1} \mathcal{Q}_{12} \rangle_{L,\alpha}
    \label{BvSoV}
    \,,\\
    \mathbb{A}_{\ell,\kappa} & =
    \mspace{45mu} \mathclap{\det_{(\gamma,j),(\delta,a)}} \mspace{48mu}
    \langle \kappa^{-i u} u^{\delta-1} \mathcal{Q}_{j}^{[a]} \rangle_{\ell,\gamma}
    \label{AlSoV}
    \,.
\end{align}
The notation is as follows:
\begin{itemize}[leftmargin=1.2em]
    \item $\mathcal{Q}_j$ and $\mathcal{Q}^k$ are normalized Q-functions:
    \begin{equation}
    \mathcal{Q}_j(u) = \frac{Q_j(u)}{\sqrt{Q_{12}(-\ii/2)}}\,, \quad \mathcal{Q}^k(u) = \frac{Q^k(u)}{\sqrt{Q^{34}(+\ii/2)}}\,.
    \end{equation}
    \item $\langle \cdot \rangle$ are contour integrals
    \begin{equation}
    \langle f\rangle_{\ell,\alpha}=\oint\frac{{\rm d}u}{(-2\pi)^\alpha}\frac{e^{2\pi u(\alpha-1)}}{(u+{i}/{2})^\ell(u-{i}/{2})^\ell}f(u)\,,
    \label{matrixelements}
\end{equation}
which encircle the poles at $\pm\ii/2$ counterclockwise.
    \item Indices run over
    \begin{align*}
    j\in\set{1,2}\,, \quad k\in\set{3,4}\,,  \qquad a,b\in\set{\pm1}
    \,, \\ \alpha,\beta\in\set{1,\dots,L}
    \,,  \qquad  \gamma,\delta\in\set{1,\dots,\ell}\,.
    \end{align*}
\item $\mathcal{N}_\ell$ is the state-independent normalization, only depending on the twists and lengths of the operators,
\begin{align}
    \mathcal{N}_\ell & = \omega^{L/2}\kappa^{\ell}\frac{(1-z_1^2)^{\frac{L}{2}}}{(1-\kappa z_1)^{\ell}}\frac{(1-z_2^2)^{\frac{L}{2}}}{(1-\kappa z_2)^{\ell}}\frac{(1-z_1z_2)^{L}}{(1-\omega z_1 z_2)^L}
    \,,
    \label{3ptNorm}
\end{align}
which is valid up to a pesky (irrelevant) phase (computed in full detail in the
Appendices) depending on the normalization of the three
operators.
\end{itemize}

An important feature of our result~\eqref{CSoV} is that it is \textit{duality invariant}~\cite{Kristjansen:2021xno}, \ie invariant under the replacement of $Q_1$, $Q_2$ with any other pair of solutions $Q_j$ of the Baxter equation. Physically, this means the structure constant is independent of which vacuum (\eg ${\rm Tr}(Z^L)$ or ${\rm Tr}(\bar{Z}^L)$) the excited state is built upon---an important consistency requirement. We prove duality invariance in \appref{app:su4sov}.

\section{Left-Right Quantum Symmetry}
\label{sec:left-right-symmetry}

Structure constants in $\mathcal{N}=4$ SYM possess an infinite tower of conservation laws beyond the usual R-charge conservation, called \emph{left-right (LR) symmetry}, which will be important in the derivation of \eqref{CSoV}. This quantum symmetry was established in \cite{Basso:2017khq} for the untwisted setup \eqref{UntwistingLimit} and it enforces $C_\ell = 0$ unless the excited state $\ket{\Psi}$ is such that $\tau_+=\tau_-$.

We will now demonstrate that in our twisted SoV formulation \eqref{CSoV}, this property is still present for generic $z_j$ and $\kappa$ once we set $\omega = 1$. Following the approach of \cite{Ekhammar:2023iph,Caetano:2020dyp,Cavaglia:2021mft}, consider a state $A$ and its image under the map
$A\rightarrow \bar{A}$, where $\bar{A}$ is obtained from $A$ by taking $Q^{{A}}_j = Q^{5-j}_{\bar{A}}$ and $\tau^{\bar{A}}_\pm = \tau^A_\mp$.
Then the Baxter operators satisfy $\mathcal{B}_{{A}}^\dagger =
\mathcal{B}_{\bar{A}}$ and we obtain
\begin{equation}
  \langle Q_2 (\mathcal{B}-\mathcal{B}^\dagger) Q_1\rangle_{L,\alpha} = 0
\end{equation}
where we dropped the state index as now all objects refer to state $A$. By shifting contours appropriately together with the QQ-relation $Q_{12} = Q_1^+ Q_2^- - Q_1^- Q_2^+$ we obtain the linear system
\begin{equation}
    \sum_{\beta=1}^L \langle u^{\beta-1}Q_{12} \rangle_{L,\alpha}(I_{+,\beta}-I_{-,\beta})=0,\quad \alpha=1,\dots,L\,,
\end{equation}
where $I_{+,\beta}$ and $I_{-,\beta}$ are the integrals of motion generated by $\tau_\pm$. For a state which is not LR symmetric at least one of the differences $I_{+,\beta}-I_{-,\beta}$ is non-zero and so the determinant of the linear system must vanish
\begin{equation}
    \det_{1\leq \alpha,\beta\leq L}\langle u^{\beta-1} Q^A_{12}\rangle_{L,\alpha} \propto \delta_{A\bar{A}}\,.
\end{equation}
Hence the structure constant~\eqref{CSoV} vanishes unless the excited state $|\Psi\rangle$ is LR symmetric.

\section{Derivation}
\label{sec:derivation}

In the following, we outline the derivation of our main
result~\eqref{CSoV}. All details of the derivation can be found in the
Appendices \ref{appReviewSU4} through \ref{AppUntwist}.

\paragraph{SoV Bases.}

The structure constant~\eqref{twistedOverlap2} is computed from
spin chain overlaps of transfer matrix eigenstates. To convert these to SoV expressions, we resolve
these overlaps in terms of left and right SoV basis states
$\bra{\svx}$ and $\ket{\svy}$, which are
generated by acting with products of transfer matrices $t_\pm(u)$, $t_0(u)$ on
sufficiently generic left and right SoV vacuum states $\bra{0}$ and
$\ket{0}$, respectively~\cite{Ryan:2018fyo}.
In this basis, the wave functions $\braket{\svx}{\Psi}$ and
$\braket{\Psi}{\svy}$ of left and right transfer matrix eigenstates
$\bra{\Psi}$ and $\ket{\Psi}$ completely factorize~\cite{Ryan:2018fyo}. An overlap $\braket{\Psi_A}{\Psi_B}$ can then be computed by inserting
$1=\sum_{\svx,\svy}\mathcal{M}_{\svx,\svy}\ket{\svy}\bra{\svx}$, where
$\mathcal{M}_{\svx,\svy}$ is obtained by inverting the matrix of
$\braket{\svx}{\svy}$ overlaps. We circumvent that by employing the
Functional Separation of Variables method (FSoV)
\cite{Cavaglia:2019pow,Gromov:2020fwh}.

\paragraph{Overlaps and Norm.}

The starting point \cite{Cavaglia:2019pow,Cavaglia:2021mft} is the observation that the operator
$\mathcal{B}^\dagger$ of the dual Baxter equation
$\mathcal{B}^\dagger Q^j=0$ satisfied by the Hodge dual Q-function
$Q^j$ is adjoint to $\mathcal{B}$,
\begin{equation}
\db{f \mathcal{B}^\dagger g}_\alpha
= \db{g \mathcal{B} f}_\alpha
\,,
\end{equation}
under the scalar product defined via
\begin{equation}
\db{f}_\alpha = \oint \frac{\dd{u}}{(-2\pi)^\alpha}
\frac{\ee^{2\pi u(\alpha-1)}}{Q_\theta^{[-2]}(u) \, Q_\theta(u) \, Q_\theta^{[2]}(u)}
f(u) \,,
\end{equation}
where the contour encircles all zeros of the denominator factors,
$Q_\theta(u)\equiv \prod_{\alpha=1}^{L}(u-\theta_\alpha)$, and
we have introduced inhomogeneities $\theta_\alpha$, $\alpha=1,\dots,L$
that shift the local Lax operators $\mathcal{L}^{(\alpha)}(u)$ used to construct the conserved charges (see Appendix \ref{appTransfer}) as $\mathcal{L}^{(\alpha)}(u) \to
\mathcal{L}^{(\alpha)}(u-\theta_\alpha)$; they are required for
regularization, and are set to $\theta_\alpha=0$ at the end.
This allows us to write
$0
= \db{Q^j_A (\mathcal{B}_A - \mathcal{B}_B) Q_k^B}_\alpha$ for any two
states $A$ and $B$ with their associated Baxter operators
$\mathcal{B}_A$ and $\mathcal{B}_B$. Since the transfer matrix eigenvalues generate integrals of motion, this equation can be
expanded to
\begin{equation}
0 = \sum_{\beta=1}^L \sum_{a,b=\pm1}
\dab{Q_A^{j,[a]} u^{\beta-1} Q_k^{B,[b]}}_\alpha
I^{AB}_{a,b,\beta}
\,,
\end{equation}
where the bracket $\dab{\cdot}_\alpha$ is obtained from
$\db{\cdot}_\alpha$ by replacing
${Q_\theta^{[-2]}Q_\theta Q_\theta^{[2]}}$ with
$Q_\theta^- Q_\theta^+$. The $I^{AB}_{a,b,\beta}$ form a complete set of
$4L$ independent differences of integrals of motion for the two
states~$A$ and~$B$. Since the vector of $I^{AB}$ is non-zero for
different states, this implies $\det \mathbb{M}_{AB}\propto\delta_{AB}$, with
\begin{equation}
   \mathbb{M}_{AB}=
   \brk[s]^2{
   \langle Q_B^{j,[a]} u^{\beta-1}Q^{A[b]}_k\rangle_{L,\alpha}
   }_{(\alpha,j,k),(\beta,a,b)}
   \,.
\end{equation}
Here we have already taken the homogeneous limit $\theta_\alpha\to0$, which
reduces $\dab{\cdot}_\alpha$ to the bracket $\avg{\cdot}$ in~\eqref{matrixelements}.
The orthogonal pairing $\det \mathbb{M}_{AB}$ must equal
$\braket{\Psi_A}{\Psi_B}$, up to an overall factor. We fix the latter by
computing the brackets $\dab{\cdot}_\alpha$ by residues, and
identifying the occurring products of factorized wave functions
$\braket{\Psi_A}{\svx}\braket{\svy}{\Psi_B}$ in the determinant, finally
arriving at
\begin{align}
\braket{\Psi_A}{\Psi_B}=
\brk^3{\frac{\mathcal{N}}{Q^A_{12}(-\ii/2)^2 \, Q_B^{34}(+\ii/2)^2}}^L
\det \mathbb{M}_{AB}
\,,
\label{eq:InnerProduct}
\end{align}
with $\mathcal{N}=(z_{13} z_{23} z_{14} z_{24})^{-1}$, $z_{ij}\equiv
z_i-z_j$. To compute the norm, we must
note that in our SoV setup, $\braket{\Psi}{\Psi}$ equals $\norm{\Psi}^2$ only up
to an overall normalization factor, see~\eqref{eqn:normratio}.
Up to a complex phase, the result is
$\norm{\Psi}^2=
|p_{13}|^{2(K-\bar{K})}|p_{14}|^{2\bar{K}}|p_{23}|^{2(M-K)}
\mathcal{N}_L \mathbb{B}$.

\paragraph{Structure Constant.}

To compute the overlap $\braket{\mathcal{W}_{\omega,\kappa}^\ell
\otimes \mathcal{W}_\omega^{L-\ell}}{\Psi}$, we first construct
$\bra{\mathcal{W}^L_\omega}$
with $\bra{\mathcal{W}_\omega}$ as in~\eqref{eq:WAdef}, where
$\mathbb{E}_{jk}$ are generators of $\alg{su}(4)$, and $p_{jk}^\omega$
have four degrees of freedom that parametrize a general global $\grp{SU}(4)$ rotation.

When $z_1z_4=z_2z_3=1$, the planar $\superN=4$ SYM structure constant obeys
$C_\ell=0$ unless $\ket{\Psi}$ is \emph{left-right symmetric}, that is
$\brk{t_+-t_-}\ket{\Psi}=0$, see \secref{sec:left-right-symmetry} above.
Thus it is sufficient
to consider BPS states $\bra{\mathcal{W}_\omega^L}$ with
$\bra{\mathcal{W}_\omega^L}\brk{t_+-t_-}=0$. These are exactly the
states with $\omega=1$,
where the parameters $p_{jk}^{\omega=1}$ retain three degrees of
freedom.
Keeping $\omega$ general, $\bra{\mathcal{W}_\omega^L}$ is a completely
general BPS state, $\bra{\mathcal{W}_\omega^L}\brk{t_+-t_-}\neq0$,
but remarkably, our result~\eqref{CSoV} still remains correct.
By a similar analysis as above, we find the overlap
\begin{equation}
\braket{\mathcal{W}_\omega^L}{\Psi} =
\frac{\omega^{L/2}}{(1-\omega z_1 z_2)^L}
\det_{\alpha,\beta} \, \avg{\omega^{-iu} u^{\beta-1} \mathcal{Q}_{12}}_{L,\alpha}
\,.
\label{eq:WPsiOverlap}
\end{equation}
Next, we deform $\bra{\mathcal{W}_\omega^{L}}$ to
$\bra{\mathcal{W}_{\omega,\kappa}^\ell \otimes
\mathcal{W}_\omega^{L-\ell}}$ by using a crucial
feature that we call \emph{localization}: By direct computation, we
find that for any global rotation
$\bra{\mathcal{W}_\omega^L}$ of $\bra{Z^L}$, the following action of
transfer matrices localizes to the first $\ell$ sites of the $L$-fold product:
\begin{equation}
\bra{\mathcal{W}_\omega^L}
\prod_{k=1}^\ell \hat{t}_{s_k}(\theta_k)
=
\bra{\mathcal{W}_\omega^\ell}
\prod_{k=1}^\ell \hat{t}_{s_k}^{(\ell)}(\theta_k)
\otimes
\bra{\mathcal{W}_\omega^{L-\ell}}
\,,
\label{eq:localized}
\end{equation}
where $s_k\in\set{+,0,-}$, $\hat{t}_{s_k}$ are properly normalized and
shifted transfer matrices, and $\hat{t}_s^{(\ell)}$ are the transfer
matrices of an $\ell$-site chain. We want to consider linear
combinations of~\eqref{eq:localized} where the first $\ell$ sites equal
$\bra{\mathcal{W}_\omega^\ell}K$, with $K\in\grp{GL}(4)$ a global
rotation.
By considering $\ell=1$, it is easy to see that there is a
one-parameter family of solutions
$\bra{\mathcal{W}_{\omega,\kappa}^\ell} = \bra{\mathcal{W}_{\omega}^\ell}K_\kappa$.
$K_\kappa$ is a polynomial of transfer matrices $t_\pm$, $t_0$ that are
diagonalized by $\ket{\Psi}$, hence
\begin{equation}
\braket{\mathcal{W}_{\omega,\kappa}^\ell \otimes \mathcal{W}_\omega^{L-\ell}}{\Psi}
= K_\kappa(\tau) \,
\braket{\mathcal{W}_\omega^{L}}{\Psi}
\,.
\end{equation}
Again considering $\ell=1$, we
recognize $K_\kappa(\tau)$ as the known SoV determinant of the
anti-fundamental $\alg{su}(3)$ spin chain~\cite{Gromov:2019wmz,Gromov:2020fwh}, which straightforwardly
generalizes to arbitrary~$\ell$~\eqref{AlSoV},
\begin{equation}
K_\kappa(\tau)
= \frac{\kappa^\ell}{(1-\kappa z_1)^\ell(1-\kappa z_2)^\ell}
\det_{(\gamma,j),(\delta,a)} \avg{\kappa^{-iu}
u^{\delta-1}\mathcal{Q}_j^{[a]}}_{\ell,\gamma}.
\label{eq:Kkappa}
\end{equation}
Combining~\eqref{eq:InnerProduct},~\eqref{eq:WPsiOverlap},
and~\eqref{eq:Kkappa}, we arrive at~\eqref{eqn:Ccovariance}, \eqref{CSoV}, where the various
prefactors combine to~\eqref{3ptNorm}, up to a complex phase that
depends on the normalization of $\ket{\Psi}$, and is spelled out in \appref{app:FullResult}.
In particular,
\eqref{eq:InnerProduct}, \eqref{eq:WPsiOverlap}, and \eqref{eq:Kkappa}
are obtained from~\eqref{eqn:sovnorm}, \eqref{eq:LRomegadet}, and~\eqref{eq:overlapHomogeneous}.

\section{Comparison with Hexagons}

The hexagon formalism \cite{Basso:2015zoa} is a well-established framework for computing three-point functions in planar $\mathcal{N}=4$ SYM, which has proven remarkably successful for structure constants of highest-weight states. This untwisted $\alg{su}(4)$ setting \cite{Basso:2017khq} is recovered from our setup in the untwisting limit \eqref{UntwistingLimit}. Highest-weight states are characterized
by sets of \textit{Bethe roots} $\textbf{v}$, $\textbf{u}$, and
$\textbf{w}$ associated to the three nodes of the $\alg{su}(4)$ Dynkin
diagram:
\begin{equation*}
    \begin{tikzpicture}[baseline={(current bounding box.center)}]
        \node[circle,draw,minimum size=5mm,inner sep=0pt,outer sep=0pt] (vnode) at (0.0,0) {};
        \node[circle,draw,minimum size=5mm,inner sep=0pt,outer sep=0pt] (unode) at (1.2,0) {};
        \node[circle,draw,minimum size=5mm,inner sep=0pt,outer sep=0pt] (wnode) at (2.4,0) {};
        \draw (vnode) -- (unode) -- (wnode);
        \node[font=\normalsize] at (0.0,0) {$\mathbf{v}$};
        \node[font=\normalsize] at (1.2,0) {$\mathbf{u}$};
        \node[font=\normalsize] at (2.4,0) {$\mathbf{w}$};
    \end{tikzpicture}\,.
\end{equation*}
The structure constants are given by
\begin{align}
    C^{\text{hex}}_{\ell}  &= \frac{\langle \textbf{v}|\textbf{w}\rangle }{\sqrt{\langle \textbf{u}|\textbf{u}\rangle }}\mathcal{A}_\ell
    \label{CHexagons}\,, \\
    \mathcal{A}_{\ell} &= \sum_{\alpha \cup \bar{\alpha} = \mathbf{u}}
    (-\tau)^{|\alpha|}\prod_{j \in \alpha}f(u_j)e^{-i p(u_j)\ell}
    \prod_{i \in \alpha, j \in \bar{\alpha}}\frac{1}{h(u_i,u_j)}\,,
    \nonumber
\end{align}
where $\langle \textbf{u}|\textbf{u}\rangle$ is the \textit{Gaudin
norm}, $\langle \textbf{v}|\textbf{w}\rangle$ is the \textit{Wing
Gaudin norm}, $\mathcal{A}_\ell$ is a sum over distributions of
the roots $\mathbf{u}$ onto the two hexagon form factors and $\tau=1$ was fixed in the original untwisted setting~\cite{Basso:2017khq}. Precise
definitions are given in \appref{AppHexagons}.

Notice that the form
of our SoV result~\eqref{CSoV} looks very similar to~\eqref{CHexagons},
with three building blocks appearing in both expressions. Remarkably,
we also get an exact match between~\eqref{CSoV} and~\eqref{CHexagons}
for $\omega=1$ but general $\kappa$ and $z_j$ by setting
$\tau = \kappa z_2$ in $\mathcal{A}_\ell$, and keeping both Gaudin
norms unchanged.
Indeed the match of the approaches works at the level of the
individual building blocks!
Namely, by setting
\begin{equation}\label{eqn:hexgauge}
    p_{13} = 1\,, \quad
    p_{14} = \frac{1-\kappa z_1}{1-z_1^2}\,,\quad
    p_{23} = \frac{1-\kappa z_2}{1-z_2^2}
    \,,
\end{equation}
we find that $\eval{C_\ell(p_{jk})}_{\omega=1}=C_\ell^{\mathrm{hex}}$, with
\begin{align}
    \langle \textbf{u}|\textbf{u}\rangle & =
    \prod\limits_{j=1}^{3}\prod\limits_{k=j+1}^{4}z_{j,k}^{\lambda_j-\lambda_k}
    \times \mathbb{B}
    \,,
    \label{MapGaudin}\\
    \langle \textbf{v}|\textbf{w}\rangle & =
    \prod\limits_{j=2}^{3}z_{1,j}^{\lambda_1 - \lambda_j}
    \times \mathbb{W}_{\omega=1}
    \,, \\
    \mathcal{A}_\ell & =
    \kappa^\ell
    \left(1-\kappa z_1\right)^{\lambda_1-\ell}
    \left(1-\kappa z_2\right)^{\lambda_2-\ell}
    \times \mathbb{A}_{\ell,\kappa} \label{MapmathcalA}
    \,.
\end{align}
Recall that the $\lambda_i$'s are related to the numbers
$K$, $M$, $\bar{K}$ of Bethe roots in the sets $\mathbf{u}$, $\mathbf{v}$,
$\mathbf{w}$ by~\eqref{eqn:dynkinroots} and~\eqref{eq:pqrlambdas}.

While the match between the norms $\mathbb{B}$ and
$\braket{\mathbf{u}}{\mathbf{u}}$ is not too surprising, it
yields interesting interpretations for the other two factors:
$\mathbb{W}_{\omega=1}$ equals the norm
$\braket{\mathbf{v}}{\mathbf{w}}$ between auxiliary Bethe roots,
whereas $\mathbb{A}_{\ell,\kappa}$ is identified as the SoV
determinant representation of the product $\mathcal{A}_\ell$ of two
hexagon form factors. We expect that the above match extends to
$\omega\neq1$ by including the corresponding $\alg{su}(4)$ rotation in the
hexagon computation.

\section{Untwisting Limits and Orbifolds}
\label{sec:untwisting}

\paragraph{Twisted theory.}

The SoV formulation \eqref{CSoV} can compute twisted structure constants at arbitrary values of $\omega$, $\kappa$ and twists $z_i$. For example, the state with $L=2$ and Dynkin labels $[1,0,1]$, detailed in \appref{AppExamples}, has structure constant (in the ``hexagon gauge'' \eqref{eqn:hexgauge})
\begin{equation}\label{eqn:twistedstructure}
   |C_{\ell=1}| = \frac{(1+\kappa)(z_2-\omega z_1)(1-z_1z_2)}{\sqrt{2z_2}(z_1-z_2)(1+z_2)(1-\omega z_1 z_2)}\,,
\end{equation}
and momentum $e^{i P} = {Q_{12}(i/2)}/{Q_{12}(-i/2)} = {1}/{z_1}$. By specializing the twists $z_k$ to
\begin{equation}\label{eqn:orbtwists}
    z_1 = e^{{2\pi i n_1}/{N}}, \quad  z_2 = e^{{2\pi i n_2}/{N}},\quad n_i\in \{ 0,1,\dots,N-1\}
\end{equation}
one obtains structure constants at the $\mathbb{Z}_N$ orbifold points of $\mathcal{N}=4$ super Yang--Mills \cite{Kachru:1998ys,Beisert:2005he,Skrzypek:2022cgg}. For states in the $\alg{su}(2)$ sub-sector, the $\mathbb{Z}_2$ orbifold tree-level structure constants were computed from hexagons \cite{lePlat:2025eod}, and all-loop Coulomb branch operator structure constants in \cite{Ferrando:2025qkr}. Our formalism reproduces those results and at the same time extends them to the full $\alg{su}(4)$ sector, yielding new predictions for general $\mathbb{Z}_N$ orbifold structure constants. For example, \eqref{eqn:twistedstructure} with \eqref{eqn:orbtwists} reduces, with $\omega=1$ and $\kappa=e^{{2 \pi i n}/{N}}$ for simplicity, to
\begin{equation}
    |C_{\ell=1}| = \frac{1}{\sqrt{2}}\frac{\cos(\pi n/N)}{\cos\left({\pi n_2}/{N}\right)}\,.
\end{equation}

\paragraph{Taking the untwisting limit.}

To recover structure constants of untwisted $\mathcal{N} = 4$ SYM, we take the limit~\eqref{UntwistingLimit}. However, this limit is subtle. When $z_i \to 1$, entire lines in the matrices \eqref{BuSoV}--\eqref{AlSoV} vanish. These zeros are compensated by the twist prefactor \eqref{3ptNorm}, which diverges as $z_i \to 1$. The combination \eqref{CSoV} is finite, and results in the untwisted structure constant.

The untwisted limit can be taken analytically in the $\alg{su}(2)$ subsector, see \appref{AppUntwist}. The vanishing determinants are controlled by a single matrix element, reducing the determinants to single minors.

For the full $\alg{su}(4)$ sector, this simplification no longer
occurs: several matrix elements vanish at the same
rate, and the determinant decomposes into a state-dependent sum of minors rather than a single minor. Only in the presence of twists does the SoV representation collapse to a single determinant for higher-rank sectors.

To extract untwisted structure constants, we opt for a simple strategy: We parametrize $z_k=e^{i \epsilon \phi_k}$, expand the Q-functions around the $\epsilon \to 0$ limit, and plug their expansions into \eqref{BuSoV}, \eqref{BvSoV} and \eqref{AlSoV}. The leading term of \eqref{CSoV} in the small twist expansion is finite, and results in the untwisted structure constant. This procedure applies equally to both primaries and descendants, as detailed in \appref{AppExamples}.

\paragraph{Counting untwisted structure constants.}

After parametrizing the twists as $z_k=e^{i \epsilon \phi_k}$ and
taking $\epsilon\rightarrow 0$, the global rotations acting on the BPS
states still retain their dependence on $\phi_k$, while the excited
state becomes independent of $\phi_k$. Accounting for $z_1 z_2 z_3
z_4=1$, we thus have five parameters: $\phi_{1}$, $\phi_{2}$,
$\phi_{3}$, $\omega$, and $\kappa$. Two generic BPS states are
parametrized by eight free parameters. However, we have three Cartan
conservation laws from the $\alg{u}(1)^3$ residual R-symmetry. Hence,
the number of parameters matches the number of degrees of
freedom. By expanding in each of these parameters, we are able to
systematically extract the structure constant of \emph{any}
excited state $|\Psi\rangle$, be it a conformal primary or a
descendant, and \emph{any} BPS states, be they rotated vacua or
descendants.

\section{Conclusion}

We took an important step towards a Q-system formulation of general
structure constants in planar $\superN=4$ SYM.
Our result~\eqref{eqn:Ccovariance} for the
$\alg{su}(4)$ scalar structure constant is in particular valid for any
values of the twists $z_i$ and polarization parameters $\omega$ and $\kappa$.
Setting these to appropriate values, we recover the structure constants of
twisted $\Integers_N$ orbifold points. Notably, we find a
one-to-one correspondence between our SoV
determinants and the building blocks of the hexagon formalism.
This direct match hints at a potential unification of the
all-loop results of~\cite{Basso:2022nny,Basso:2025mca} with SoV.

Our formalism may be adapted to
settings where Hexagons are not
fully developed, such as ABJM theory \cite{Pereira:2017unx}, where
recent progress on orthogonal group Q-systems can be exploited
\cite{Ekhammar:2021myw}, or marginal deformations of planar
$\mathcal{N}=4$ SYM \cite{Cavaglia:2020hdb,Eden:2022ipm}.

Structurally, the twists $z_j$ as well as $\omega$ and $\kappa$ play a similar
role as the cusp angles in the Maldacena--Wilson loop
\cite{Polyakov:1980ca,Gromov:2015dfa}.
Strikingly, our result indeed takes the same form as the finite
coupling results of~\cite{Cavaglia:2018lxi,Cavaglia:2021mft}, where
each inner product is a pairing between the excited state Q-functions
and twisted vacuum Q-functions represented by the factors
$\omega^{-iu}$ and $\kappa^{-iu}$ in \eqref{BvSoV} and \eqref{AlSoV}.

Our result is the ideal starting point for
loop corrections, where the Q-functions get promoted to their higher-loop QSC counterparts. It also
enables analytic continuation in spin, and direct access to light-ray
operators \cite{Homrich:2024nwc}, bypassing the integer-spin
restriction inherent in the hexagon picture.
Finally, the functional SoV
techniques developed here pave the way towards a supersymmetric SoV
formalism for the full $\alPSU(2,2|4)$ superalgebra at finite coupling
-- the ultimate goal of this program.

\section*{Acknowledgments}

We are grateful to S.~Ekhammar, N.~Gromov, C.~Kristjansen, A.~Liashyk, T.~Skrzypek, and P.~Vieira
for discussions. T.\,B.\ and P.\,R.\ are grateful to M.~Baroni and V.~Schomerus for discussion and collaboration on related topics.
A preliminary version of these results was presented
at ``New Frontiers of Quantum Field and Gravity" in Beijing in January
2026. P.\,R.~is grateful to the organizers and participants, and
especially B.~Basso, Y.~Jiang and D.~Serban for many stimulating
discussions related to this work.
The work of C.\,B. was supported by the European Research
Council (ERC) under the European Union's Horizon 2020 research and
innovation program - 60 - (grant agreement No.~865075) EXACTC.
The work of T.\,B. and P.\,R. was funded by the Deutsche Forschungsgemeinschaft (DFG, German Research Foundation) -- 460391856.
T.\,B. and P.\,R. acknowledge support from DESY (Hamburg, Germany), a member of the Helmholtz Association HGF,
and by the Deutsche Forschungsgemeinschaft
(DFG, German Research Foundation) under Germany's Excellence Strategy
– EXC 2121 ``Quantum Universe'' –- 390833306.

\appendix

\ifarxiv
\protected\long\def\preto#1#2{%
  \edef#1{%
    \unexpanded{#2}%
    \unexpanded\expandafter{#1}%
  }%
}
\preto\section{%
\ifnum\value{section}=14\addtocounter{section}{1}\fi%
\ifnum\value{section}=8\addtocounter{section}{2}\fi%
}
\fi

\section{Review of the \texorpdfstring{$\alg{su}(4)$}{su(4)} Sector}
\label{appReviewSU4}

The $\alg{su}(4)$ sector is built out of the six complex scalar fields
$Z,\bar{Z},X,\bar{X},Y,\bar{Y}$. The six fields define a
basis of the anti-symmetric representation $\mathbb{C}^4\wedge\mathbb{C}^4$ of the global $\alg{su}(4)$ R-symmetry algebra with Cartan weights
\begin{align*}
    Z&:[0,\phantom{+}1,0]\,, & X&:[\phantom{+}1,-1,\phantom{+}1]\,, & Y&:[\phantom{+}1,0,-1]\,, \nn \\
    \bar{Z}&:[0,-1,0]\,, & \bar{X}&:[-1,\phantom{+}1,-1]\,,&\bar{Y}&:[-1,0,\phantom{+}1]\,.
\end{align*}
It is often convenient to extend this algebra by an additional $\alg{u}(1)$
charge and introduce the $\alg{gl}(4)$ algebra, spanned by generators
$\mathbb{E}_{ij}$ subject to the commutation relations
\begin{equation}
    [\mathbb{E}_{ij},\mathbb{E}_{kl}]=\delta_{jk}\mathbb{E}_{il}-\delta_{li}\mathbb{E}_{kj}\,.
    \label{eq:GL4comm}
\end{equation}

The $\alg{su}(4)$ Cartan generators $\mathbb{H}_k$ are then given by
$\mathbb{H}_k = \mathbb{E}_{kk}-\mathbb{E}_{k+1,k+1}$.

\paragraph{Building the representation.}

An explicit realization of the 6-dimensional representation can be constructed as follows. Let $e_{jk}$ denote the usual $4\times 4$ unit matrices with $1$ in position $(j,k)$ and $0$ everywhere else, and let $e_j$ denote unit basis vectors in $\mathbb{C}^4$ with $1$ in position $j$ and $0$ everywhere else. Then a basis $v_1,\dots,v_6$ of $\mathbb{C}^4\wedge \mathbb{C}^4$ is given by
\begin{equation}
    \begin{split}
        & v_1 = e_1 \wedge e_2,\quad v_2 = e_1\wedge e_3,\quad v_3 = e_1 \wedge e_4, \\
        & v_4 = e_2\wedge e_3,\quad v_5 = e_2\wedge e_4,\quad v_6 = e_3\wedge e_4,
    \end{split}
\end{equation}
where $v\wedge w = (v\otimes w - w \otimes v)/\sqrt{2}$. Then the action of $\alg{gl}(4)$ on this space is simply given by
\begin{equation}
    \mathbb{E}_{jk}(v\wedge w) = (e_{jk}v)\wedge w + v\wedge(e_{jk}w)\,.
\end{equation}
The Lie algebra generators defined in this way possess natural Hermitian conjugation properties $\mathbb{E}_{ij}^\dagger = \mathbb{E}_{ji}$ where $\dagger$ denotes transpose and complex conjugation.

\paragraph{Relating fields and spin chain states.}

By matching their Cartan weights, we identify the six complex scalar fields with the basis vectors $v_j$ by
\begin{equation}\label{eqn:fieldmap}
    \{|Z\rangle,|X\rangle ,|Y\rangle,|\bar{Y}\rangle,|\bar{X}\rangle,|\bar{Z}\rangle \}\sim \{v_1,v_2,v_3,v_4,v_5,v_6\}\,.
\end{equation}
The $Z$ vector is the highest-weight state of the
representation: $\mathbb{E}_{jk}|Z\rangle =0$, $j<k$, with
$\mathfrak{gl}(4)$ highest weights
$[\nu_1,\nu_2,\nu_3,\nu_4]=[1,1,0,0]$, \ie
$\mathbb{E}_{jj}|Z\rangle =\nu_j |Z\rangle$ We fix the relative normalization in \eqref{eqn:fieldmap} by imposing
\begin{equation}
\begin{split}
    & |X\rangle = -\mathbb{E}_{32}|Z\rangle,\quad |Y\rangle = \mathbb{E}_{42} |Z\rangle ,\quad |\bar{Y}\rangle = \mathbb{E}_{31}|Z\rangle \\
    & |\bar{X}\rangle = \mathbb{E}_{41}|Z\rangle ,\quad |\bar{Z}\rangle = \mathbb{E}_{41}\mathbb{E}_{32}|Z\rangle\,.
\end{split}
\end{equation}
Left states $\langle Z|, \langle X|, \dots, \bra{\bar{Z}}$ are then defined by orthonormality, $\langle Z|Z\rangle=1$, $\langle Z|\bar{Z}\rangle =0$ \etc.

\paragraph{Overlaps vs Wick contractions.}

On the field theory side, Wick contractions are only non-zero between
a field and it's conjugate, \eg
\begin{equation}
\overset{\raisebox{-0.3ex}{\rotatebox{90}{$]$}}}{Z\bar{Z}}=1
\quad \text{and} \quad
\overset{\raisebox{-0.3ex}{\rotatebox{90}{$]$}}}{ZZ}=0
\,.
\end{equation}
On the other hand, in the spin chain language the overlaps are $\langle Z|Z\rangle=1$, $ \langle \bar Z|Z\rangle=0$, \etc. In order to associate fields with spin chain states so that spin chain overlaps are consistently identified with Wick contractions, we follow \cite{Escobedo:2010xs} and make the association
\begin{align}
    & Z \mapsto |Z\rangle,\quad \bar{Z} \mapsto |\bar{Z}\rangle, \quad \text{for ket states}\,, \\
    & Z \mapsto \langle \bar{Z}|,\quad \bar{Z} \mapsto \langle Z|, \quad \text{for bra states}\,,
\end{align}
with all other fields mapped similarly. Note that despite the fact that this identification swaps a field and its conjugate for bra states, there is no complex conjugation of coefficients, and so for example
\begin{equation}
    e^{i \phi} Z\mapsto e^{i\phi}\langle \bar{Z}|
\end{equation}
for real $\phi$. Additionally, the BPS state $\langle Z|e^{\mathbb{E}_{14}+\mathbb{E}_{23}}$ is given in terms of basis fields by
\begin{equation}
    \langle Z|e^{\mathbb{E}_{14}+\mathbb{E}_{23}} =  \langle Z| + \langle\bar{Z}| - \langle X| + \langle\bar{X}|\,,
\end{equation}
which maps to the linear combination of fields $Z+\bar{Z}-\bar{X}+X.$

In this way, the fields associated to the vectors in \eqref{eq:WAdef} take the explicit form
\begin{multline}
    W_\omega =\bar{Z}+\nu_1 p^{\omega}_{23} \bar{X} +\nu_2 p_{24}^\omega \bar{Y}-\nu_2 p^\omega_{23} Y \\
    -\nu_1 p^\omega_{14} X +(\nu_2^2 p^\omega_{13}p^\omega_{24} - \nu_1^2 p^\omega_{14}p^\omega_{23}) Z
    \,,
\end{multline}
where
\begin{equation}\label{eqn:nuconstraints}
    \nu_1 = \left(\frac{z_{13}z_{24}}{z_{12}z_{34}} \right)^{1/2},\quad \nu_2 = \left(\frac{z_{23}z_{14}}{z_{12}z_{34}}\right)^{1/2},\quad z_{ij}:=z_i-z_j\,.
\end{equation}
Finally, the polarizations $p^\omega$ and $p^{\omega,\kappa}$ in \eqref{eq:WAdef} are related to the ``reference" polarizations $p_{ij}$ by
\begin{equation}
\begin{split}
& p^\omega_{23} = p_{23}\frac{1 - \omega z_1 z_3}{1 - \omega z_1 z_2}, \quad p^\omega_{24} = p_{24}\frac{1 - \omega z_1 z_4}{1 - \omega z_1 z_2}\,,\\
& p^\omega_{13} = p_{13}\frac{1 - \omega z_2 z_3}{1 - \omega z_1 z_2}, \quad p^\omega_{14} = p_{14}\frac{1 - \omega z_2 z_4}{1 - \omega z_1 z_2}\,,
\end{split}
\end{equation}
and
\begin{equation}
p^{\omega,\kappa}_{ij} = p^{\omega}_{ij} \frac{1-\kappa z_j}{1-\kappa z_i}\,.
\end{equation}

\section{Transfer Matrices}
\label{appTransfer}

In the following, we explicitly build the complete set of commuting integrals of motion whose eigenstates are the twisted operators entering the twisted structure constant.

\paragraph{Lax and monodromy matrices.}

The key ingredient is the Lax operator $\mathcal{L}_{jk}(u)$ defined by
\begin{equation}
    \mathcal{L}_{jk}(u) = u\delta_{jk}-i\, \mathbb{E}_{kj},\quad j,k=1,2,3,4\,.
    \label{eq:LaxOp}
\end{equation}
The Hilbert space of the spin chain will be $(\mathbb{C}^6)^{\otimes
L}$, and we denote the Lax operator acting on the $\alpha$'th copy of
$\mathbb{C}^6$ as $\mathcal{L}_{jk}^{(\alpha)}(u)$. From here we build
the twisted monodromy matrix $T_{jk}(u)$ with
\begin{equation}
    T_{jk}(u) = z_j \sum_{\mathclap{i_1,\dots,i_{L-1}}}\mathcal{L}^{(1)}_{j i_1}(u-\theta_1)\dots \mathcal{L}^{(L)}_{i_{L-1}k}(u-\theta_L)
    \,,
\end{equation}
where the indices $i_\alpha$ are summed from $1$ to $4$. The parameters $\theta_\alpha$ are inhomogeneities -- useful regulators in the operatorial SoV construction, which will be put to $0$ in the structure constant.

The twist parameters $z_j$ are independent, but we can impose $z_1 z_2 z_3 z_4=1$ by a trivial rescaling. In the remaining sections, we will fix $z_4$ by imposing $z_1 z_2 z_3 z_4=1$, although sometimes we will display $z_4$ explicitly if it leads to simpler expressions.

By standard arguments \cite{Minahan:2002ve} the transfer matrix $t_+(u) = \sum_{j=1}^4 T_{jj}(u)$ generates a commuting family of integrals of motion $\hat{I}_{+,\beta}$, $\beta=1,\dots,L$
\begin{equation}
    t_+(u) = \chi_+ u^L + \sum_{\beta=1}^L u^{\beta-1}\hat{I}_{+,\beta},\quad
    [t_+(u),t_+(v)]=0
    \,,
    \label{eq:tplusIOM}
\end{equation}
where $\chi_+=z_1+z_2+z_3+z_4$. In particular the global $\alg{su}(4)$
Cartan generators $\mathcal{E}_{jj}-\mathcal{E}_{j+1,j+1}$, $\mathcal{E}_{jj} = \sum_{\alpha=1}^L \mathbb{E}_{jj}^{(\alpha)}$ are encoded in $\hat{I}_{+,L}$ with
\begin{equation}
    \hat{I}_{+,L}= -i\sum_{j=1}^4 z_j \mathcal{E}_{jj} - \sum_{\alpha=1}^L \theta_\alpha\,.
\end{equation}

\paragraph{Fusion.}

In addition to $t_+(u)$ we can build more transfer matrices and integrals of motion using the fusion procedure \cite{Zabrodin:1996vm}. For this we use antisymmetric combinations of~$p$ monodromy matrices
\begin{align}
   &T\left[^{j_1 \dots j_p}_{k_1\dots k_p}\right](u) \\ \nonumber
   & \mspace{20mu} =\sum_{\sigma \in\grp{S}_p}(-1)^{|\sigma|}T_{j_{\sigma(1)}k_1}(u)T_{j_{\sigma(2)}k_2}^{[-2]}(u)\dots T_{j_{\sigma(p)}k_p}^{[-2(p-1)]}(u)
   \,,
\end{align}
where we used the shorthand notation $f^{[a]}(u) = f(u+\tfrac{a i}{2})$ for shifts of the spectral parameter and $\grp{S}_p$ denotes the permutation group on $p$ letters. With these, the additional transfer matrices $t_0(u)$ and $t_-(u)$ are given by
\begin{align}
    t_0(u) &= \sum_{1\leq j_1<j_2 \leq 4} T\left[^{j_1 j_2}_{j_1 j_2}\right](u+\tfrac{i}{2})
    \,,\\
    Q_\theta^{[2]}(u)Q_\theta^{[-4]}(u)t_-(u) &=
    \mspace{-10mu} \sum_{1\leq j_1<j_2<j_3 \leq 4} \mspace{-10mu}
    T\left[^{j_1 j_2 j_3}_{j_1 j_2 j_3}\right](u+i)
    \,,
\end{align}
where $Q_\theta(u)=\prod_{\alpha=1}^L(u-\theta_\alpha)$
encode some trivial overall zeros which originate from the fusion procedure.

The fused transfer matrices
$t_0(u)$ and $t_-(u)$ generate an additional $2L$ and $L$ integrals of motion respectively
\begin{equation}
\begin{split}
    & t_0(u) = \chi_0 u^{2L}+\sum_{\beta=1}^{2L} u^{\beta-1} \hat{I}_{0,\beta} \,, \\
    & t_-(u) = \chi_- u^{L}+\sum_{\beta=1}^{L} u^{\beta-1} \hat{I}_{-,\beta}\,,
\end{split}
\label{eq:t0minusIOM}
\end{equation}
where $\chi_0$ and $\chi_-$ denote some simple combination of twist parameters,
\begin{equation}
    \begin{split}
       & \chi_0 = z_1 z_2 + z_1 z_3 +z_1 z_4 +z_2 z_3 +z_2 z_4+z_3 z_4\,, \\
       & \chi_- = \frac{1}{z_1}+\frac{1}{z_2} + \frac{1}{z_3} + \frac{1}{z_4}\,.
    \end{split}
\end{equation}

All charges generated by $t_\pm(u)$ and $t_0(u)$ mutually commute, and
furthermore form a complete set, \ie there are no additional
independent charges which commute with these ones \cite{Ryan:2020rfk}. Two particularly important charges are $t_0(0)$ and $t_0(0)^{-1}\partial_u t_0(0)$. In the homogeneous limit $\theta_\alpha\rightarrow 0$, these
correspond to the discrete shift (momentum) operator around the chain, as well as the twisted one-loop dilatation operator $D\propto \sum_{j=1}^L \mathcal{H}_{j,j+1}$ of planar $\mathcal{N}=4$ SYM in the $\alg{su}(4)$ sector \cite{Minahan:2002ve}
\begin{equation}
    \mathcal{H}_{j,j+1} = 2 + 2 P_{j,j+1} - K_{j,j+1}
    \,,
\end{equation}
where $P_{j,j+1}$ is the permutation operator on sites $j,j+1$, and $K_{j,j+1}$ is the trace operator.
Additionally, the joint spectrum of the $4L$ charge operators is non-degenerate, as long as the twist parameters $z_j$ are all distinct.

\paragraph{Hermitian conjugation.}

Thanks to the Hermitian conjugation property $\mathbb{E}_{ij}^\dagger = \mathbb{E}_{ji}$ of the $\mathfrak{gl}(4)$ generators one can show the following Hermitian conjugation properties of the transfer matrices
\begin{equation}
    t_+(u)^\dagger = t_-(u^*+i),\quad t_0(u)^\dagger = t_0(u^*+i)\,,
\end{equation}
assuming the twist eigenvalues $z_j$ are pure phases $z_j = e^{i\phi_j}$ and the inhomogeneities $\theta_\alpha$ are real. This ensures that the set of conserved charges is closed under Hermitian conjugation.

\paragraph{Left and right eigenvectors.}

Joint eigenvectors of the mutually commuting transfer matrices $t_\pm(u)$ and $\tau_0(u)$ are denoted $|\Psi\rangle$ and their eigenvalues $\tau_\pm(u)$ and $\tau_0(u)$ respectively:
\begin{equation}
    t_\pm(u) |\Psi\rangle = \tau_\pm(u) |\Psi\rangle,\quad t_0(u) |\Psi\rangle = \tau_0(u) |\Psi\rangle\,.
\end{equation}
In addition to right eigenvectors $|\Psi\rangle$ we also have left eigenvectors $\langle \Psi|$ corresponding to the same eigenvalue as~$|\Psi\rangle$,
\begin{equation}
   \langle \Psi| t_\pm(u) = \tau_\pm(u) \langle\Psi|,\quad \langle \Psi|t_0(u)= \tau_0(u) \langle \Psi|\,.
\end{equation}
We do not yet enforce any relative normalization between left and right eigenvectors. In particular, we do not assume they are Hermitian conjugates. To make this distinction clear, we will denote the Hermitian conjugate $|\Psi\rangle^\dagger$ of $|\Psi\rangle$ as $\langle \Psi^*|$ and stress that in general $\langle \Psi|\neq \langle \Psi^*|$. Thanks to the non-degeneracy of the transfer matrix spectrum, these two vectors differ only by a normalization. We will later fix this normalization in a way which is more useful from the SoV point of view.

\section{Baxter equation and Q-functions}
\label{appSU4Baxter}

The eigenvalues $\tau_\pm(u)$ and $\tau_0(u)$ of the transfer matrices can be elegantly encoded in functional equations -- Baxter TQ equations and QQ-relations.

\paragraph{Baxter equation.}

The Baxter equation is a fourth-order finite-difference equation involving the finite-difference operator
\begin{align}\label{eqn:baxter}
  \mathcal{B} = Q_\theta^{[2]} Q_\theta D^{-4}
  &- \tau_+ Q_\theta^{[2]}D^{-2}
  \nonumber \\
  + \tau_0^+
  &- \tau_-^{[2]} Q_\theta^{[-2]}D^{2}
  + Q_\theta^{[-2]}Q_\theta D^{4}
  \,,
\end{align}
where $D$ denotes the shift operator which acts on functions as $Df(u)
= f(u+\tfrac{i}{2})$, and we use the notation $f^\pm:=f(u\pm\tfrac{i}{2})$. The solutions to the Baxter equation
$\mathcal{B}Q_j=0$, $j=1,\dots,4$, are \emph{Q-functions}. In the
compact $\alg{su}(4)$ sector, they are twisted polynomials
\begin{equation}
    Q_j(u) =z_j^{-i u} \times q_j(u)
\end{equation}
where $q_j(u)$ are polynomials of degree $\lambda_j$
\begin{equation}
q_j(u)\simeq A_j u^{\lambda_j}+\dots \,.
\label{eq:qAsymp}
\end{equation}
The exponents $\lambda_j$ encode the $\alg{su}(4)$ Dynkin labels $[p,q,r]$ of the excited state  via
\begin{align}
    p = \lambda_2-\lambda_1,\quad q = \lambda_3-\lambda_2,\quad  r = \lambda_4-\lambda_3\,
\end{align}
and the constraint $2L =\lambda_1+\lambda_2+\lambda_3+\lambda_4$.

\paragraph{QQ-relations.}

The $Q_j$ functions can be embedded into a zoo of Q-functions called Q-system. These are a collection of functions $Q_A(u)$ where $A\subset\{1,2,3,4\}$ with $Q_\emptyset(u):=1$. Q-functions with more than one index can be constructed as quantum Wronskians of the $Q_j$. For functions $f_1,\dots,f_n$, the quantum Wronskian $W(f_1,\dots, f_n)$ is defined by
\begin{equation}
    W(f_1,\dots, f_n) = \det_{1\leq i,j\leq n} f_i^{[n+1-2j]}\,.
\end{equation}
For example, $W(f_1,f_2) = f_1^+ f_2^- - f_1^- f_2^+$\,. Then we define the additional Q-functions $Q_{ij}$ and $Q_{ijk}$ by the \emph{QQ-relations}
\begin{gather}
Q_{ij} = W(Q_i,Q_j)
\,, \quad
Q_\theta Q_{ijk} = W(Q_i,Q_j,Q_k)
\,, \nn \\
Q_{1234} = W(Q_1,Q_2,Q_3,Q_4)
\,,
\label{Q12WronksDef}
\end{gather}
which are clearly anti-symmetric in their indices. Furthermore, we demand that all $Q_A(u)$ are twisted polynomials
\begin{equation}
    Q_A =  q_A(u)\prod_{a\in A} z_a^{-i u}
\end{equation}
where $q_A(u)$ is a polynomial.

\paragraph{Q-function normalization.}

The normalization constants $A_j$ entering the large-$u$ behavior~\eqref{eq:qAsymp} of the Q-functions $Q_j$ will propagate to all other Q-functions by the QQ-relations. Physical quantities (\eg conformal dimensions) only depend on ratios of Q-functions and so this normalization is unphysical. We will fix all normalization freedom by imposing that $q_1, q_{12}, q_{123}$ and $q_{1234}$ are monic polynomials. We will assume this normalization in what follows.

\paragraph{Quantization condition.}

The Baxter equation implies the constraint
\begin{equation}
  Q_{1234}(u) =  Q_\theta^+(u) Q_\theta^-(u)
  \label{eq:QQrelation}
\end{equation}
which follows from the zero determinant
\begin{equation}
    \left|\begin{array}{ccccc}
        Q_j^{[-4]} & Q_j^{[-2]} & Q_j & Q_j^{[2]} & Q_j^{[4]}  \\
        Q_1^{[-4]} & Q_1^{[-2]} & Q_1 & Q_1^{[2]} & Q_1^{[4]}  \\
        Q_2^{[-4]} & Q_2^{[-2]} & Q_2 & Q_2^{[2]} & Q_2^{[4]}  \\
        Q_3^{[-4]} & Q_3^{[-2]} & Q_3 & Q_3^{[2]} & Q_3^{[4]}  \\
        Q_4^{[-4]} & Q_4^{[-2]} & Q_4 & Q_4^{[2]} & Q_4^{[4]}  \\
    \end{array} \right|=0\,.
\end{equation}
Clearly this relation must be equivalent to the Baxter equation, as
cofactor expansion yields a fourth-order finite-difference equation on
$Q_j$ which vanishes for $j=1,2,3,4$. By comparing the coefficients of
$Q_j^{[-4]}$ and $Q_j^{[4]}$ with~\eqref{eqn:baxter}, we obtain
\eqref{eq:QQrelation} up to an overall multiplicative constant, which
equals~$1$ in our normalization conventions.
Conversely, the Baxter equation~\eqref{eqn:baxter} follows
from~\eqref{eq:QQrelation}, with the transfer matrix eigenvalues
$\tau_\pm$, $\tau_0$ given in~\eqref{eqn:teigenvalues} below.

The Q-system provides an alternative approach to computing the spectrum of conserved charges, which is often substantially more convenient than directly solving the Baxter equation or Bethe equations \cite{Marboe:2016yyn}. By solving these relations with the specified analytic properties, we obtain a one-to-one correspondence between solutions and joint eigenstates of the transfer matrices, with the Q-functions encoding the Bethe roots $\mathbf{v}=\{ v_1,\dots,v_K\}, \mathbf{u}=\{u_1,\dots,u_M \}, \mathbf{w}=\{w_1,\dots,w_{\bar{K}}\}$ of the state entering the Hexagon expression \eqref{CHexagons} according to
\begin{gather}
   q_1(u) = \prod_{j=1}^K (u-v_j)
   \,,\quad
   q_{123}(u) = \prod_{j=1}^{\bar{K}} (u-w_j)
   \,, \nn \\
   q_{12}(u) = \prod_{j=1}^M (u-u_j)\,.
\end{gather}

\paragraph{Transfer matrix eigenvalues.}

The Baxter equation together with the QQ-relations allows us write the transfer matrix eigenvalues directly in terms of Q-functions. For this it is useful to introduce the combinations $\Lambda_j(u)$ with
\begin{align}
  \Lambda_1 & = Q_\theta^{[-2]}\frac{Q_1^{[2]}}{Q_1}
  \,, &
  \Lambda_2 & = Q_\theta^{[-2]}\frac{Q_1^{[-2]}}{Q_1}\frac{Q_{12}^{+}}{Q_{12}^{-}}
  \,, \nn \\
  \Lambda_3 & = Q_\theta \frac{Q_{12}^{[-3]}}{Q_{12}^-}\frac{Q_{123}}{Q_{123}^{[-2]}}
  \,, &
  \Lambda_4 & = Q_\theta \frac{Q_{123}^{[-4]}}{Q_{123}^{[-2]}}\,.
\label{eqn:quantumeigenvalues}
\end{align}
Then one can show that by setting
\begin{align}
\label{eqn:teigenvalues}
    \tau_+ &= \Lambda_1+\Lambda_2+\Lambda_3+\Lambda_4 \\ \nn
    \tau_0 &= \Lambda_4^+ \Lambda_3^- \!+ \Lambda_4^+ \Lambda_2^- \!+\Lambda_4^+ \Lambda_1^- \!+\Lambda_3^+ \Lambda_2^- \!+ \Lambda_3^+ \Lambda_1^- \!+\Lambda_2^+ \Lambda_1^- ,
\end{align}
and by setting $\tau_-$ to the same expression as $\tau_+$ but with
$Q_1$ interchanged with $Q_{123}$ then~\eqref{eq:QQrelation} implies the Baxter equation~\eqref{eqn:baxter}.

\paragraph{Hodge-dual Q-functions.}

The Q-functions $Q_{ijk}$ with three indices are on essentially equal footing with the single-index Q-functions $Q_j$, and it is convenient to denote them as Q-functions with upper indices $Q^j$ with
\begin{equation}
    Q^1 = -Q_{234},\ Q^2=Q_{134},\ Q^3 = -Q_{124},\ Q^4=Q_{123}\,.
\end{equation}
More generally, for any subset $A\subset \{ 1,2,3,4\}$ we define $Q^A$ by
\begin{equation}
    Q^A(u) = \varepsilon^{\bar{A} A}Q_{\bar{A}}(u),\quad \varepsilon^{1234}=1\,,
\end{equation}
where $\bar{A}$ denotes the complement of $A$ in $\{1,2,3,4\}$. These Q-functions are called \textit{Hodge dual Q-functions}. Similar to the functions $Q_j$, the $Q^j$ functions satisfy a \emph{dual Baxter equation} $\mathcal{B}^\dagger Q^j=0$ with
\begin{align}\label{eqn:dualbaxter}
\mathcal{B}^\dagger =
  Q_\theta^{[2]} Q_\theta D^{-4}
  &- \tau_- Q_\theta^{[2]}D^{-2}
  \nonumber \\
  + \tau_0^{+}
  &- \tau_+^{[2]} Q_\theta^{[-2]}D^{2}
  + Q_\theta^{[-2]}Q_\theta D^{4}
  \,.
\end{align}
The Baxter operator \eqref{eqn:baxter} and dual Baxter operator \eqref{eqn:dualbaxter} will play an important role in what follows.

Finally, QQ-relations for the Hodge dual Q-functions can be deduced from the QQ-relations of the original Q-functions, in particular we have the relation
\begin{equation}
    W(Q^j,Q^k) = Q^{jk}\,.
\end{equation}

\section{SoV Wave Functions}\label{app:su4sov}

We will now describe separation of variables for the transfer matrix eigenstates. The separation of variables for the $\alg{su}(4)$ spin chain was carried out in \cite{Ryan:2018fyo,Ryan:2020rfk}. It is heavily based on representation theory, and a key aspect of the construction is to label states by combinatorial objects called Gelfand--Tsetlin patterns, which we now review.

\paragraph{Gelfand--Tsetlin patterns.}

For our six-dimen\-sion\-al representation, with
$\mathfrak{gl}(4)$ weights $[\nu_1,\nu_2,\nu_3,\nu_4]=[1,1,0,0]$, Gelfand--Tsetlin patterns are arrays of the following shape:
\begin{equation}\label{eqn:GTpattern}
    \begin{array}{cc@{\,}c@{}c@{}c@{\,}cc}
        1 & \ & 1 & \ & 0 & \ & 0 \\
         \ & 1 & \ & \mu_{12} & \ & 0 \\
         \ & \ & \mu_{11} & \ & \mu_{22} \\
         \ & \ & \ & \mu_{21}
    \end{array}\,,
\end{equation}
where $\mu_{ij}\in \{0,1\}$ must satisfy the constraints
\begin{equation}
    \mu_{11}\geq \mu_{21}\geq \mu_{22},\quad \mu_{11}\geq \mu_{12}\geq \mu_{22}\,.
\end{equation}
It is easy to check that there are six such patterns that can be drawn, matching the number of states. For a chain of length $L$, each site $\alpha$ comes with its own patterns with coordinates $\mu_{ij}^\alpha$. We will use this labeling in what follows.

\paragraph{Building the left SoV basis \texorpdfstring{$\langle \svx|$}{<x|}.}

We now explain how to build the SoV basis $\langle \svx|$, following \cite{Ryan:2018fyo} which substantially expands on the initial idea of \cite{Maillet:2018bim} proposed for the fundamental representation.

The starting point is a certain reference state that we denote
$\langle 0|$. It will be called the SoV vacuum. It should be a generic
enough vector so that repeated action of conserved charges on it
generates a basis of our Hilbert space. We will take it to be a global
$\grp{GL}(4)$ rotation
acting on the Z-vacuum $\langle Z^L|$. A
convenient parametrization is as follows
\begin{equation}
    \langle 0| = \langle Z^L|e^{V_{13}\mathbb{E}_{13}}e^{V_{14}\mathbb{E}_{14}}e^{V_{23}\mathbb{E}_{23}}e^{V_{24}\mathbb{E}_{24}}
    \,,
    \label{eq:SOVvacuumDef}
\end{equation}
where $V_{ij}$ are some complex numbers, which for the moment are free. The SoV basis, parametrized by the $4L$ numbers $\mu^\alpha_{kj}$, is then given by
\begin{equation}\label{eqn:xbasis}
   \langle \svx|= \langle 0|\prod_{\alpha=1}^L t_{\mu^\alpha_1}(\theta_\alpha)t_{\mu^\alpha_2}(\theta_\alpha)
\end{equation}
where $\mu_1^\alpha$ and $\mu_2^\alpha$ denote Young diagrams $\mu_j^\alpha=[\mu_{j1}^\alpha,\mu_{j2}^\alpha]$
with $t_{[0,0]}(u)=1$ and
\begin{equation}\label{eqn:TsGT}
    t_{[1,0]}(u) = \frac{t_+(u)}{Q_\theta^{[-2]}(u)}
    \,, \quad
    t_{[1,1]}(u)=\frac{t_0^-(u)}{Q_\theta^{[-2]}(u)Q_\theta^{[-4]}(u)}\,.
\end{equation}
In particular, the SoV vacuum state $\bra{0}$~\eqref{eq:SOVvacuumDef}
has Gelfand--Tsetlin pattern $\mu_{jk}^\alpha=0$ on all
sites~$\alpha$.

The separated variables for our spin chain are a collection of $4L$
operators $\hat{\svx}^\alpha_{jk}$, $\alpha=1,\dots L$, $j,k\in
\{1,2\}$ which can be constructed from the B-operator formalism \cite{Gromov:2016itr}. By definition, they are diagonal in the SoV basis $\langle \svx|$,
\begin{equation}
    \langle \svx|\hat{\svx}^\alpha_{jk} = \svx^\alpha_{jk}\langle \svx|
    \,,
    \label{eq:xDef}
\end{equation}
with eigenvalues
$\svx^\alpha_{jk}=\theta_\alpha+i(\mu^\alpha_{jk}-k+1)$, where $\mu_{jk}^\alpha$ are nodes on the Gelfand--Tsetlin pattern \eqref{eqn:GTpattern} \cite{Ryan:2018fyo}. By construction, these are a complete set of commuting operators, since all Gelfand--Tsetlin labels can be trivially reconstructed from the
eigenvalues.
The fact that these operators $\hat{\mathsf{x}}^\alpha_{jk}$ are diagonal in the basis defined by~\eqref{eqn:xbasis} is non-trivial, and we
refer the reader to~\cite{Ryan:2018fyo} for the proof.

\paragraph{Right SoV wave functions.}

The wave function $\Psi(\svx)$ of conserved charge eigenstates $|\Psi\rangle$ in the basis~\eqref{eqn:xbasis} is given by~\cite{Ryan:2018fyo}
\begin{equation}\label{eqn:xwavefunction}
    \Psi(\svx)=\langle \svx|\Psi\rangle = \prod_{\alpha=1}^L \frac{\psi(\svx^\alpha_{11},\svx^\alpha_{12})}{Q_{12}(\theta_\alpha-\tfrac{i}{2})}\frac{\psi(\svx^\alpha_{21},\svx^\alpha_{22})}{Q_{12}(\theta_\alpha-\tfrac{i}{2})}
    \,,
\end{equation}
where
\begin{equation}\label{eqn:onepartwave}
    \psi(\svx_1,\svx_2)=Q_1(\svx_1)Q_2(\svx_2)-Q_2(\svx_1)Q_1(\svx_2)
    \,,
\end{equation}
and we have conveniently normalized the transfer matrix eigenstates
$|\Psi\rangle$ by imposing $\langle 0|\Psi\rangle =1$.

The form of the wave function~\eqref{eqn:xwavefunction} follows
from~\eqref{eqn:xbasis} by noting that
the eigenvalues $\tau_{[1,0]}(u)$ and $\tau_{[1,1]}(u)$ of $t_{[1,0]}(u)$ and $t_{[1,1]}(u)$ satisfy
\begin{align}\label{eqn:tauwave}
    \tau_{[1,0]}(\theta_\alpha) &= \frac{Q_{[1}(\theta_\alpha + i) Q_{2]}(\theta_\alpha-i)}{Q_{12}(\theta_\alpha - \frac{i}{2})} = \frac{\psi(\theta_\alpha+i,\theta_\alpha-i)}{Q_{12}(\theta_\alpha - \frac{i}{2})} \\
    \tau_{[1,1]}(\theta_\alpha) &= \frac{Q_{[1}(\theta_\alpha + i) Q_{2]}(\theta_\alpha)}{Q_{12}(\theta_\alpha - \frac{i}{2})} = \frac{\psi(\theta_\alpha+i,\theta_\alpha)}{Q_{12}(\theta_\alpha - \frac{i}{2})}
    \,,
\end{align}
which is an immediate consequence of \eqref{eqn:teigenvalues} and the QQ-relations \eqref{Q12WronksDef}.

We see that the wave function is completely factorized across the sites $\alpha$, justifying the term ``separated variables'', and is furthermore factorized across the ``dual'' diagonals $[\mu_{11}^\alpha,\mu_{12}^\alpha]$ and $[\mu_{21}^\alpha,\mu^\alpha_{22}]$ of the Gelfand--Tsetlin pattern~\eqref{eqn:GTpattern}.

\paragraph{Duality transformations.}

In the above, we expressed the SoV wave functions
$\braket{\svx}{\Psi}$ in terms of $Q_1$, $Q_2$, and $Q_{12}$.
As mentioned in the main text, there is nothing special about this
choice, and we could just as well have used any other combination of~$Q_j$, $Q_k$, and $Q_{jk}$.

The appearance of $Q_1$, $Q_2$, and $Q_{12}$ in the SoV wave function
ultimately stems from writing the transfer matrix eigenvalues in terms
of the objects $\Lambda_j$ \eqref{eqn:quantumeigenvalues}.
These objects specify a \textit{nesting path}, a path in the Q-system
from $Q_\emptyset$ to $Q_{1234}$ obtained by adding one additional index at each step, in particular we have used
\begin{equation}
    Q_\emptyset\rightarrow Q_1 \rightarrow Q_{12} \rightarrow Q_{123} \rightarrow Q_{1234}\,.
\end{equation}
We could just as well have used another path, \eg
\begin{equation}
    Q_\emptyset \rightarrow Q_2\rightarrow Q_{24}\rightarrow Q_{124}\rightarrow Q_{1234}
    \,,
\end{equation}
then \eqref{eqn:quantumeigenvalues} would be replaced with
\begin{align}
  \Lambda_1 & = Q_\theta^{[-2]}\frac{Q_2^{[2]}}{Q_2}
  \,, &
  \Lambda_2 & = Q_\theta^{[-2]}\frac{Q_2^{[-2]}}{Q_2}\frac{Q_{24}^{+}}{Q_{24}^{-}}
  \,, \nn \\
  \Lambda_3 & = Q_\theta \frac{Q_{24}^{[-3]}}{Q_{24}^-}\frac{Q_{124}}{Q_{124}^{[-2]}}
  \,, &
  \Lambda_4 & = Q_\theta \frac{Q_{124}^{[-4]}}{Q_{124}^{[-2]}}\,.
\label{eqn:quantumeigenvaluesDual}
\end{align}
Such changes of nesting path are called \emph{duality
transformations}. While individual $\Lambda_j$ are \emph{not}
invariant under duality transformations, the full combinations
entering the transfer matrix eigenvalues \eqref{eqn:teigenvalues} are
invariant, and hence so are the relations \eqref{eqn:tauwave} and
hence so are the SoV wave functions \eqref{eqn:xwavefunction}. The wave functions in each of the additional SoV bases we construct below are also duality invariant. All of
the objects entering the twisted structure constant in the main text
are built from SoV wave functions and hence are also invariant under
duality transformations. See \cite{Kazakov:2015efa,Ryan:2020rfk} for
further discussion.

\paragraph{Right SoV basis \texorpdfstring{$|\svy\rangle$}{|y>}.}

We now build the dual SoV basis which we denote as $|\svy\rangle$. It is designed to factorize the left transfer matrix eigenstates $\langle \Psi|$ with the following wave functions
\begin{equation}\label{eqn:ywave}
    \Psi(\svy)=\langle \Psi|\svy\rangle  = \prod_{\alpha=1}^L \frac{\psi(\svy^\alpha_{11},\svy^\alpha_{12})}{Q^{34}(\theta_\alpha+\tfrac{i}{2})}\frac{\psi(\svy^\alpha_{21},\svy^\alpha_{22})}{Q^{34}(\theta_\alpha+\tfrac{i}{2})}
\end{equation}
with $\svy^\alpha_{kj}=\theta_\alpha +i(\mu^\alpha_{kj}-j+1)$.

To build it explicitly, we again need a reference state, this time denoted as $|1\rangle$ which we take to be
\begin{equation}\label{eqn:rightsovvac}
    |1 \rangle = e^{W_{42}\mathbb{E}_{42}}e^{W_{41}\mathbb{E}_{41}} e^{W_{32}\mathbb{E}_{32}}e^{W_{31}\mathbb{E}_{31}}|Z^L\rangle \,.
\end{equation}
We use the notation $|1\rangle$ as this state will correspond to Gelfand--Tsetlin patterns with all $\mu^\alpha_{kj}=1$, and we normalize the left eigenstates $\langle \Psi|$ according to $\langle \Psi|1\rangle=1$. We also need some convenient combinations of transfer matrices. This time we define $\bar{t}_{[1,1]}(u)=1$ and
\begin{equation}
    \bar{t}_{[1,0]}(u)=\frac{t_+^{[2]}(u)}{Q_\theta^{[2]}(u)},\quad \bar{t}_{[0,0]}(u) =\frac{t_0^{[4]}(u)}{Q_\theta^{[2]}(u)Q_\theta^{[4]}(u)}\,.
\end{equation}
Then we can build the basis explicitly as
\begin{equation}\label{eqn:ybarbasis}
    |\bar{\svy}\rangle = \prod_{\alpha=1}^L \bar{t}_{\mu^\alpha_1}(\theta_\alpha)\bar{t}_{\mu_2^\alpha}(\theta_\alpha)|1\rangle\,.
\end{equation}
Using the QQ-relations and Baxter equation we can check that \eqref{eqn:ybarbasis} indeed implies \eqref{eqn:ywave}. Note that we can define $|0\rangle$, the state with all $\mu^\alpha_{kj}=0$, as
\begin{equation}
    |0\rangle = \prod_{\alpha=1}^L \bar{t}_{[0,0]}(\theta_\alpha)\bar{t}_{[0,0]}(\theta_\alpha)|1\rangle\,.
\end{equation}

\paragraph{SoV measure.}

The scalar product between two transfer matrix eigenstates $\langle
\Psi_A|\Psi_B\rangle$ can be computed by inserting the resolution of
identity $1=\sum_{\svx,\svy}|\svy\rangle\langle \svx|
\mathcal{M}_{\svx,\svy}$ in the SoV bases. In principle, the measure
$\lM_{\svx,\svy}$ can be obtained from the inverse matrix of overlaps
$\langle \svx|\svy\rangle$. In practice however, the measure is a very complicated object~\cite{Gromov:2020fwh} and depends on the parameters $V_{ij}$ and $W_{ij}$ entering the two reference states \eqref{eq:SOVvacuumDef} and \eqref{eqn:rightsovvac}. It is highly advantageous to impose that the $\langle\svx|$ and $|\svy\rangle$ bases are related by a certain anti-automorphism of the underlying algebraic structure, see \cite{Gromov:2020fwh,Ryan:2021duf}. It can be imposed by demanding  that $\langle \svx|0\rangle \propto \delta_{\svx,0}$ and $\langle 0|\svy\rangle \propto \delta_{0,\svy}$.%
\footnote{This is related to both $\langle\svx|$ and $|\svy\rangle$ diagonalizing the so-called SoV charge operator of \cite{Gromov:2019wmz,Gromov:2020fwh}, which measures the number of SoV excitations above the SoV vacuum states.}
To resolve these constraints it is convenient to introduce $p_{ij}$ with
\begin{equation}\label{eqn:Vconstraints}
 V_{13} = \nu_2 p_{13}, \ V_{23} = \nu_1 p_{23},\ V_{14} = \nu_1 p_{14},\  V_{24} = \nu_2 p_{24}
 \,,
\end{equation}
with $W_{ji}$ defined similarly with $p_{ij}\rightarrow p_{ji}$. The parameters $\nu_1$ and $\nu_2$ depend on the twists $z_j$ according to~\eqref{eqn:nuconstraints}.
With this parametrization of $V_{ij}$ and $W_{ji}$, the constraints simply read $p_{ji}p_{ij}=(-1)^{i+j}$ and $p_{13}p_{24}=p_{23}p_{14}$. From now on, we will assume these constraints have been satisfied.

In practice, while we could compute the measure directly, it is much easier to avoid computing $\mathcal{M}_{\svx,\svy}$ and instead compute $\langle \Psi_A|\Psi_B\rangle$ directly using the Functional Separation of
Variables method that we now outline.

\section{SoV Scalar Product}

\paragraph{Orthogonal product.}

We now construct the SoV scalar product using the functional approach. The starting point is the family of integrals $\db{\cdot}_\alpha$ defined for (twisted) polynomial functions~$f(u)$ by
\begin{equation}
    \db{f}_\alpha = \oint\frac{{\rm d}u}{(-2\pi)^\alpha} \frac{e^{2\pi u(\alpha-1)}}{Q_\theta^{[-2]}Q_\theta Q_\theta^{[2]}}f(u)
    \,,
\end{equation}
with the contour encircling all poles of the integrand at $\theta_\alpha,\theta_\alpha\pm i$, $\alpha=1,\dots,L$. Their key property is that the Baxter operator $\mathcal{B}$ \eqref{eqn:baxter} and dual Baxter operator $\mathcal{B}^\dagger$ \eqref{eqn:dualbaxter} are adjoint to each other
\begin{equation}\label{eqn:Baxadjoint}
    \db{f\mathcal{B}^\dagger g}_\alpha = \db{g\mathcal{B} f}_\alpha\,.
\end{equation}
Now we consider the difference $\db{Q^j_A(\mathcal{B}_A-\mathcal{B}_B)Q^B_k}_\alpha=0$ for two states $A$ and $B$, which vanishes thanks to the adjointness condition \eqref{eqn:Baxadjoint}. Plugging in the explicit form of the Baxter operator~\eqref{eqn:baxter}, we obtain
\begin{multline}
  \db^1{Q_A^j\tau_{+,AB}^{[2]}Q_\theta^{[-2]}Q_k^{B[2]}}_\alpha
  + \db^1{Q_A^j\tau_{-,AB}Q_\theta^{[2]}Q_k^{B[-2]}}_\alpha \\
  - \db^1{Q_A^j\tau_{0,AB}^{+}Q_k^{B}}_\alpha=0
  \,,
\end{multline}
where we denoted $\tau_{AB} = \tau_A-\tau_B$, \ie the differences of transfer matrices for the two different states. Now expand the transfer matrices into a basis of $4L$ integrals of motion. A convenient choice is
\begin{equation}
    \tau^{[2]}_{+,AB} = \sum_{\beta=1}^L (u+\tfrac{i}{2})^{\beta-1}I^{AB}_{+,\beta},\quad \tau_{-,AB} = \sum_{\beta=1}^L (u-\tfrac{i}{2})^{\beta-1}I^{AB}_{-,\beta}
\end{equation}
and
\begin{equation}
\begin{split}
    \tau_{0,AB}^{+}&= Q_\theta^{[-2]}\sum_{\beta=1}^L (u+\tfrac{i}{2})^{\beta-1}I^{AB}_{0,+,\beta}\\
    & + Q_\theta^{[+2]}\sum_{\beta=1}^L (u-\tfrac{i}{2})^{\beta-1}I^{AB}_{0,-,\beta}
\end{split}
\end{equation}
where $I_{\pm,\beta}^{AB}$ and $I_{0,\pm,\beta}^{AB}$,
$\beta=1,\dots,L$ denote the differences of $4L$ integrals of motion
for two different states~$A$ and~$B$, and we introduced the combinations $I_{0,\pm,\beta}$ which are particularly convenient linear combinations of the eigenvalues of the operators $\hat{I}_{0,\beta'}$ defined in \eqref{eq:t0minusIOM}.
The convenience of this choice comes from the fact that we can now shift the contours to obtain
\begin{align}
  0 = \mathord{} & \sum_{\beta=1}^L \dab{Q_A^{j-}u^{\beta-1}Q_k^{B+}}_\alpha I_{+,\beta}^{AB}
  + \dab{Q_A^{j-}u^{\beta-1}Q_k^{B-}}_\alpha I_{0,+,\beta}^{AB}
  \nn \\
  & - \dab{Q_A^{j+}u^{\beta-1}Q_k^{B+}}_\alpha I_{0,-,\beta}^{AB}
  + \dab{Q_A^{j+}u^{\beta-1}Q_k^{B-}}_\alpha I_{-,\beta}^{AB}
  \,,
\label{eqn:linsystem}
\end{align}
where now we have the new bracket $\dab{\cdot}_\alpha$ defined by
\begin{equation}\label{eqn:inhombracket}
    \dab{f}_\alpha = \oint\frac{{\rm d}u}{(-2\pi)^\alpha} \frac{e^{2\pi u(\alpha-1)}}{Q_\theta^{-} Q_\theta^{+}}f(u),\quad \alpha=1,\dots,L\,.
\end{equation}
The equation \eqref{eqn:linsystem} for $\alpha=1,\dots,L$ and
$i,j\in\{1,2\}$ constitutes a system of $4L$ linear equations for the
$4L$ differences of integrals of motion. Since the integrals of motion
completely specify a state, at least one of these differences must be
non-zero for two different states $A$ and $B$, hence the determinant
of the linear system must satisfy
\begin{equation}
\det M_{AB}\propto \delta_{AB}
\,, \quad
M_{AB} = \dab^1{Q_A^{j[s_1]}u^{\beta-1}Q_k^{B[s_2]}}_\alpha
\,,
\label{eq:detM}
\end{equation}
with rows of $M_{AB}$ labeled by triples $(\alpha,j,k)$, and columns labeled by
$(\beta,s_1,s_2)$, with $s_i\in\{+,-\}$.

\paragraph{Wave function overlap.}

The determinant $\det M_{AB}$ constitutes an orthogonal pairing of states $A$, $B$, and therefore must equal the wave function overlap $\braket{\Psi_A}{\Psi_B}$, up to a (possibly state-dependent) overall factor. In order to make the relation precise, one can
compute the integrals~\eqref{eqn:inhombracket} by residues and take the determinant. The result will clearly be an anti-symmetric combination of $Q^B_1$ and $Q^B_2$ with various shifts, and similarly for $Q_A$. For low length~$L$, one can explicitly rearrange the terms and combine everything into the combinations $\psi(\svx^\alpha_{11},\svx^{\alpha}_{12})\psi(\svx^\alpha_{21},\svx^{\alpha}_{22})$ defined in \eqref{eqn:onepartwave}, and similarly for~$\svy$. For higher length, one needs to perform some manipulations on the determinant, similar to \cite{Gromov:2020fwh} in the $\alg{su}(3)$ setting, but the conclusion is the same.

In order to bring the sum of terms to a sum of products of SoV wave functions \eqref{eqn:xwavefunction} and \eqref{eqn:ywave}, we divide the determinant~\eqref{eq:detM} by $Q_{12}^-(\theta_\alpha)^2 Q^{34+}(\theta_\alpha)^2$. The result is then a sum of precisely the form
\begin{equation}\label{eqn:waveresol}
    \sum_{\svx,\svy}\Psi_A(\svy)\Psi_B(\svx)\mathcal{M}_{\svx,\svy}\,.
\end{equation}
Since the transfer matrix eigenstates $\langle \Psi_A|$ and $|\Psi_B\rangle$ form a basis, we can then extract the resolution of identity
\begin{equation}\label{eqn:resolutionidentity}
    1 = \sum_{\svx,\svy}|\svy\rangle \langle \svx|\mathcal{M}_{\svy,\svx}
\end{equation}
and read off the measure $\mathcal{M}_{\svy,\svx}$. We just need to fix the overall normalization of the determinant, which should be state independent as all state information coming from the resolution \eqref{eqn:resolutionidentity} is through the SoV wave functions in \eqref{eqn:waveresol}.

To fix this normalization, we can compute the determinant for some simple state. We know that $\langle Z^L|Z^L\rangle=1$ (see Appendix \ref{appReviewSU4}) and the corresponding Q-functions are
\begin{equation}
    Q_j(u) = z_j^{-i u},\quad Q^j(u) = z_j^{iu}\,.
\end{equation}
Evaluating the determinant in this instance is simple, and we can read off the required factor to reproduce $\langle Z^L|Z^L\rangle=1$. It is some Vandermonde-like expression involving the inhomogeneities $\theta_\alpha$ and the twist parameters. We will not spell this factor
out explicitly. Instead, we take the homogeneous limit $\theta_\alpha\to0$. All of the
poles at $\theta_\alpha\pm\tfrac{i}{2}$ in the integrands~\eqref{eqn:inhombracket} collapse to $\pm {i}/{2}$, but each entry of the matrix $M_{AB}$
remains finite and non-zero, and the integrals are replaced according to $\dab{\cdot}_\alpha\rightarrow \langle \cdot \rangle_{L,\alpha}$, where
\begin{equation}
    \langle f\rangle_{L,\alpha} =\oint\frac{{\rm d}u}{(-2\pi)^\alpha} \frac{e^{2\pi u(\alpha-1)}}{(u-\tfrac{i}{2})^L(u+\tfrac{i}{2})^L}f(u)
    \,,
\end{equation}
and we made the $L$-dependence explicit, as it will be useful later. The prefactor requires some care, but
at the end of the day we finally obtain
\begin{equation}\label{eqn:sovnorm}
   \langle \Psi_A|\Psi_B\rangle = \left(\frac{\mathcal{N}}{\left(Q^A_{12}(-\tfrac{i}{2})Q^{34}_B(+\tfrac{i}{2})\right)^2}\right)^L \det\mathbb{M}_{AB}
   \,,
\end{equation}
with the matrix
\begin{equation}
   \mathbb{M}_{AB}= \langle Q_B^{j,[s_1]} u^{\beta-1}Q^{A[s_2]}_k\rangle_{L,\alpha}
   \,,
\end{equation}
where rows are labeled by $(\alpha,j,k)$, and columns by $(\beta,s_1,s_2)$,
and $\mathcal{N}$ denotes the simple factor
\begin{equation}
    \mathcal{N}= (z_{13}z_{23}z_{14}z_{24})^{-1},\quad z_{ij}:=z_i-z_j\,.
    \label{eq:Nfactor}
\end{equation}

\paragraph{Relation to the norm.}

As we stressed in \appref{appTransfer}, the state $\langle \Psi|$ is not equal to the Hermitian conjugate $\langle \Psi^*|$ of $|\Psi\rangle$, and hence \eqref{eqn:sovnorm} for $A=B$ is not equal to the squared norm $||\Psi||^2:=\langle \Psi^* |\Psi\rangle$  of the state $|\Psi\rangle$.  However, we do have $\langle \Psi|\propto \langle \Psi^*|$ and the (state-dependent) conversion factor is actually rather straightforward to work out.

As discussed in the previous section, the states are normalized according to
\begin{align}\label{eqn:rightnormal}
   \langle Z^L|e^{V_{13}\mathbb{E}_{13}}e^{V_{14}\mathbb{E}_{14}}e^{V_{23}\mathbb{E}_{23}}e^{V_{24}\mathbb{E}_{24}}|\Psi\rangle &= 1
   \,, \\
\label{eqn:leftnormal}
   \langle \Psi|e^{W_{42}\mathbb{E}_{42}}e^{W_{41}\mathbb{E}_{41}} e^{W_{32}\mathbb{E}_{32}}e^{W_{31}\mathbb{E}_{31}}|Z^L\rangle &=1\,.
\end{align}
By taking the Hermitian conjugate of \eqref{eqn:rightnormal} we obtain
\begin{equation}\label{eqn:conjugatenormal}
    \langle \Psi^*|e^{V_{24}^*\mathbb{E}_{42}}e^{V_{14}^*\mathbb{E}_{41}} e^{V_{23}^*\mathbb{E}_{32}}e^{V_{13}^*\mathbb{E}_{31}}|Z^L\rangle =1\,.
\end{equation}
We can then compute the relative normalization between $\langle \Psi|$ and $\langle \Psi^*|$ by computing the ratio of \eqref{eqn:conjugatenormal} and \eqref{eqn:leftnormal}.

To do this, we use the fact that $\langle \Psi|$ and $|Z^L\rangle$ have definite Cartan quantum numbers. By expanding the exponentials, not all terms will contribute, but only those for which the overall Cartan quantum numbers are conserved between the initial and final state. The result will be a finite sum of the form
\begin{equation}
\begin{split}
   \sum_{n's} & \langle \Psi|(W_{42}\mathbb{E}_{42})^{n_{42}} (W_{41}\mathbb{E}_{41})^{n_{41}}\\
   &\times (W_{32}\mathbb{E}_{32})^{n_{32}}(W_{31}\mathbb{E}_{31})^{n_{31}}|Z^L\rangle =1
   \,,
\end{split}
\end{equation}
where the sum is over all values of $n$'s such that the global quantum numbers match. In general, there will be many terms contributing, each with various different powers of $W_{ji}$. However, the constraints we imposed on $W_{ji}$ and $V_{ij}$ in the previous section, see \eqref{eqn:Vconstraints} and \eqref{eqn:nuconstraints}, now play a crucial role. By a careful analysis of the contributing terms, one can show that once these constraints are satisfied the $W$-dependence contributes only by an overall factor depending on the Cartan quantum numbers, or equivalently the number of Bethe roots $K,M,\bar{K}$. A similar analysis can be carried out on \eqref{eqn:conjugatenormal}. Since the overall dependence on the parameters $V$ and $W$ only enters through an overall factor, we can then deduce the precise proportionality between $\langle\Psi|$ and $\langle \Psi^*|$ and hence compute the norm $||\Psi||^2 = \langle \Psi^*|\Psi\rangle$. The result is

\begin{equation}\label{eqn:normratio}
\begin{split}
   \frac{\langle \Psi^*|\Psi\rangle}{\langle \Psi|\Psi\rangle} &= \left(-\frac{\nu_1}{\nu_1^*}\right)^{K+M+\bar{K}} \left(\frac{\nu_2}{\nu_2^*}\right)^{K+\bar{K}} \\
   & \times |p_{13}|^{2(K-\bar{K})}|p_{14}|^{2\bar{K}}|p_{23}|^{2(M-K)}\,.
\end{split}
\end{equation}
By combining \eqref{eqn:normratio} with \eqref{eqn:sovnorm} we obtain
the SoV expression for the norm $||\Psi||$, which we will write in a
form similar to~\eqref{BuSoV}. Introduce
\begin{equation}
     \mathbb{B}=
    \mspace{45mu} \mathclap{\det_{(\alpha,j,k),(\beta,a,b)}} \mspace{48mu}
    \langle \mathcal{Q}_{j}^{[a]} u^{\beta-1} \mathcal{Q}^{k[b]}  \rangle_{L,\alpha}
    \,,
\end{equation}
with $\mathcal{Q}_j = Q_j / \sqrt{Q_{12}(-{\ii}/{2})}$ and $\mathcal{Q}^j = Q^j / \sqrt{Q^{34}(+{\ii}/{2})}$. Then we have
\begin{equation}\label{eqn:finalnorm}
\begin{split}
    ||\Psi||^2 = \mathord{} &
    \mathcal{N}^L
    \times\mathbb{B} \\
    & \times  \left(-\frac{\nu_1}{\nu_1^*}\right)^{K+M+\bar{K}} \left(\frac{\nu_2}{\nu_2^*}\right)^{K+\bar{K}} \\
   & \times |p_{13}|^{2(K-\bar{K})}|p_{14}|^{2\bar{K}}|p_{23}|^{2(M-K)}\,,
\end{split}
\end{equation}
with $\mathcal{N}$ as in~\eqref{eq:Nfactor}.

\section{Integrable Boundary State}

Having computed the norm $||\Psi||$ in the SoV representation we now move towards computing the overlap $\langle \mathcal{W}_{\omega,\kappa}^\ell\otimes \mathcal{W}^{L-\ell}_\omega|\Psi\rangle$. In this section we will describe how to compute this overlap for $\ell=0$, \ie $\langle \mathcal{W}_\omega^L|\Psi\rangle$.

We will proceed as follows. First, in the untwisting limit structure constants admit a left-right (LR) symmetry selection rule \cite{Basso:2017khq}. As we established in the main text, this symmetry condition can be extended to the twisted setup. We will construct an overlap which manifests this symmetry property and then deform away from the symmetric point, thus introducing an additional twist angle into the game. This will be precisely the definition of $\langle \mathcal{W}_\omega|$.

\paragraph{The LR symmetry condition.}

As we established in the main text, Functional SoV picks out a distinguished determinant of Q-functions
\begin{equation}\label{eqn:LRdet}
 \det_{1\leq \alpha,\beta\leq L} \dab{u^{\beta-1}Q_{12}}_\alpha
\end{equation}
which vanishes unless the transfer matrix eigenstate $\ket{\Psi}$ described by the Q-functions is
LR-symmetric. Here we are using the inhomogeneous brackets $\dab{f}_{\alpha}$
from \eqref{eqn:inhombracket}. We can interpret this determinant as an
overlap $\langle W|\Psi\rangle$ with a final state $\langle W|$ that is
an integrable boundary state which is independent of $\ket{\Psi}$ and preserves half of
the conserved charges, $\langle W|(t_+(u) - t_-(u))=0$.

It is easy to show that this state actually exists. All we need to do
is compute the integrals in \eqref{eqn:LRdet} by residues, expand the
determinant, and check that the whole expression can be recast as a
sum over SoV wave functions $\braket{\svx}{\Psi}$ with coefficients $W_\svx$. By comparing with the general form of
the wave functions we can confirm that this is indeed the case, and
furthermore we can read off that only two possible configurations of
SoV states contribute to $\bra{W}$. These correspond to
Gelfand--Tsetlin patterns
\begin{equation}\label{eqn:redGTpats}
        \begin{array}{ccccccc}
        1 & \ & 1 & \ & 0 & \ & 0 \\
         \ & 1 & \ & 0 & \ & 0 \\
         \ & \ & 0 & \ & 0 \\
         \ & \ & \ & 0
    \end{array}\quad \text{and} \quad    \begin{array}{ccccccc}
        1 & \ & 1 & \ & 0 & \ & 0 \\
         \ & 1 & \ & 1 & \ & 0 \\
         \ & \ & 1 & \ & 0 \\
         \ & \ & \ & 0
    \end{array}\,.
\end{equation}
We will now deduce the precise form of the state $\langle W|$.

\paragraph{The state \texorpdfstring{$\langle W_\omega|$}{<W_w|}.}

Actually, we will construct a more general state $\langle W_\omega|$,
from which the integrable boundary state $\bra{W}$ is recovered in the
$\omega\to1$ limit. In the structure constant (three-point correlator) \eqref{twistedOverlap2} that we ultimately want to compute
the parameter $\omega$ twists one of the three operators. This breaks
the LR symmetry condition: For $\omega\neq1$, the structure constant
will become non-zero even for non-LR-symmetric states $\ket{\Psi}$.
Accordingly, unlike $\bra{W}$, $\bra{W_\omega}$ does not preserve the
charges $(t_+-t_-)$, that is $\bra{W_\omega}\brk{t_+(u)-t_-(u)}\neq0$.
The state $\bra{W_\omega}$ is defined by the condition
\begin{equation}
  \braket{W_\omega}{\Psi} =
  \Theta_L\prod_{\gamma=1}^L \frac{\omega^{1/2+i\theta_\gamma}r_\omega}{Q^-_{12}(\theta_\gamma)} \det_{1\leq \alpha,\beta\leq L} \dab{u^{\beta-1} \omega^{-iu} Q_{12}}_\alpha
  \,,
  \label{eq:LRomegadet}
\end{equation}
where we have conveniently normalized the expression, with $r_\omega = (1-\omega\, z_1 z_2)^{-1}$ and $\Theta_L$ is some unimportant Vandermonde determinant-like expression of the inhomogeneities $\theta_\alpha$ which goes to $1$ in the homogeneous limit.
To compute the state, we should write $\langle W_\omega|=\sum_{\svx} w_\svx\langle \svx|$ where the sum is over the SoV states $\langle \svx|$ built from the Gelfand--Tsetlin patterns \eqref{eqn:redGTpats}. The coefficients $w_\svx$ completely specify the state $\langle W_\omega|$. They can be computed by comparing $\langle W_\omega|\Psi\rangle = \sum_{\svx}w_{\svx}\,\Psi(\svx)$ with the right-hand-side of \eqref{eq:LRomegadet}.

We did this explicitly for $L=1,2,3,4$, and found that $\langle W_\omega|$ is actually a product state $\langle W_\omega|=\langle \mathcal{W}^L_\omega|$, with
\begin{equation}
    \langle \mathcal{W}_\omega|=\langle Z| e^{R_{13}\mathbb{E}_{13}}e^{R_{14}\mathbb{E}_{14}}e^{R_{23}\mathbb{E}_{23}}e^{R_{24}\mathbb{E}_{24}}
    \label{eq:WL}
\end{equation}
where
\begin{equation}
\begin{split}
   &  \frac{R_{13}}{V_{13}} = \frac{1-\omega z_2 z_3}{1-\omega z_1 z_2},\quad  \frac{R_{14}}{V_{14}} = \frac{1-\omega z_2 z_4}{1-\omega z_1 z_2}\,. \\
   &  \frac{R_{23}}{V_{23}} = \frac{1-\omega z_1 z_3}{1-\omega z_1 z_2},\quad  \frac{R_{24}}{V_{24}} = \frac{1-\omega z_1 z_4}{1-\omega z_1 z_2}\,,
\end{split}
\end{equation}
and $V_{ij}$ are the parameters in~\eqref{eq:SOVvacuumDef}
and $R_{ij}$ is related to $p_{ij}^\omega$ appearing in \eqref{eq:WAdef} by
\begin{equation}\label{eqn:pol1}
\begin{split}
   & R_{13} = \nu_2 p^\omega_{13},\quad R_{23}=\nu_1 p^\omega_{23},\\
   & R_{14}=\nu_1 p^\omega_{14},\quad R_{24}=\nu_2 p^\omega_{24}\,.
\end{split}
\end{equation}
The fact that $\langle W_\omega|$ is a product state is actually not too surprising, and we can establish it rigorously for any length $L$. The details however require some technical machinery, namely the \textit{SoV embedding morphism} developed in \cite{Ryan:2020rfk}. We will not describe this here.

\section{The Complete Overlap}

In the previous section, we outlined where the relation~\eqref{eq:LRomegadet}
comes from, which matches the result \eqref{CSoV} announced in the main text for $\ell=0$. We will now extend this to $\ell>0$. To do this, we will describe a novel feature of the SoV basis which we refer to as \emph{SoV localization}.

\paragraph{SoV localization.}

The localization property can be stated as follows. Let $K$ denote a
global $\grp{GL}(4,\mathbb{C})$ transformation, and consider the action of
transfer matrices~\eqref{eqn:TsGT} on a rotated vacuum:
\begin{equation}
    \langle Z^L|K \prod_{\alpha=1}^\ell t_{[1,0]}(\theta_\alpha)
    \,.
\end{equation}
Remarkably, this action of transfer matrices is \textit{localized} to the first $\ell$ sites of the spin chain, and furthermore precisely matches the action of a transfer matrix built for a chain of length $\ell$. More precisely,
\begin{equation}
    \langle Z^L|K \prod_{\alpha=1}^\ell t_{[1,0]}(\theta_\alpha) =  \langle Z^L|K \prod_{\alpha=1}^\ell t_{[1,0]}^{(\ell)}(\theta_\alpha)
    \,,
\label{eqn:localization}
\end{equation}
where
\begin{equation}
t_{[1,0]}^{(\ell)}(u) = \frac{t_+^{(\ell)}(u)}{Q_{\theta,\ell}^{[-2]}(u)}
\,,\quad
Q_{\theta,\ell}(u) = \prod_{\alpha=1}^\ell (u-\theta_\alpha)
\,,
\end{equation}
and $t_+^{(\ell)}(u)$ denotes the transfer matrix for the length $\ell$ spin chain, which can be written in terms of Lax operators as
\begin{equation}
    t_+^{(\ell)}(u) = \mathcal{L}_{i_1 i_2}(u-\theta_1)\dots \mathcal{L}_{i_\ell i_1}(u-\theta_\ell)z_{i_1}\,.
\end{equation}
The same localization property~\eqref{eqn:localization} holds for
$t_{[1,1]}(u)$~\eqref{eqn:TsGT}, which is built from the fused transfer
matrix $t_0(u)$. Since all transfer matrices commute, the same is true
for mixed products of fused and unfused transfer matrices.
The proof of the localization property~\eqref{eqn:localization} is
rather technical, so we postpone it to \appref{sec:localization}. For
now, we will examine its consequence.

As a warm-up, we will consider the case $\ell=1$. Denote $\langle K^L|=\langle Z^L|K$ and consider for instance the action of localizing transfer matrices
\begin{equation}\label{eqn:loc1}
    \langle K^L|(1 -\kappa\, t_{[1,0]}(\theta_1)+\kappa^2 t_{[1,1]}(\theta_1))\,.
\end{equation}
By the localization property, we know that this will be a state of the form $\langle \mathcal{K}_\kappa \otimes K^{L-1}|$ for some
$\mathcal{K}_\kappa\in \Complex^4 \wedge \Complex^4$.
On the other hand, we can also write
\begin{equation}\label{eqn:loc2}
    \langle \mathcal{K}_\kappa\otimes K^{L-1}|\Psi\rangle = \langle K^L|\Psi\rangle(1 -\kappa\, \tau_{[1,0]}(\theta_1)+\kappa^2 \tau_{[1,1]}(\theta_1))
\end{equation}
since $|\Psi\rangle$ diagonalizes the transfer matrices in \eqref{eqn:loc1}. By direct computation, we can express this combination of transfer matrix eigenvalues in \eqref{eqn:loc2} as the determinant
\begin{equation}\label{eqn:Adet1}
  \frac{\kappa^{1+2i\theta_1}}{Q_{12}^-(\theta_1)}  \left|\begin{array}{cc}
        \dab{\kappa^{-i u}Q_1^+}_{1,1} & \dab{\kappa^{-i u}Q_1^-}_{1,1}  \\
        \dab{\kappa^{-i u}Q_2^+}_{1,1} & \dab{\kappa^{-i u}Q_2^-}_{1,1}
    \end{array} \right|\,,
\end{equation}
where
\begin{equation}\label{eqn:inhombracketell}
    \dab{f}_{\ell,\alpha}
    = \oint\frac{\dd{u}}{(-2\pi)^\alpha} \frac{e^{2\pi u(\alpha-1)}}{Q_{\theta,\ell}^{-} \, Q_{\theta,\ell}^{+}}f(u)
    \,, \quad
    \alpha=1,\dots,\ell
    \,.
\end{equation}
is the scalar bracket~\eqref{eqn:inhombracket} restricted to $\ell$ sites.
This determinant is actually a familiar object \cite{Gromov:2019wmz,Gromov:2020fwh} and corresponds to the SoV scalar product but for a spin chain in the anti-fundamental representation of $\alg{su}(3)$. The upshot of this is that from \cite{Gromov:2019wmz,Gromov:2020fwh} we immediately know how to generalize this expression to any $\ell$, it is simply
\begin{equation}\label{eqn:loc3}
   \prod_{\gamma=1}^\ell \frac{\kappa^{1+2i\theta_\gamma}}{Q_{12}^-(\theta_\gamma)} \det_{(a,\alpha),(b,\beta)} \dab{\kappa^{-iu} u^{\beta-1}Q_a^{[b]}}_{\ell,\alpha}\,,
\end{equation}
where the rows and columns of the matrix are labeled by pairs $(a,\alpha)$ and $(b,\beta)$ respectively, with the pairs lexicographically ordered with
\begin{align}
   (a,\alpha)& \in \{(1,1),\dots,(1,\ell),(2,1),\dots,(2,\ell) \} \nn \\
   (b,\beta) & \in \{(1,1),\dots,(1,\ell),(-1,1),\dots,(-1,\ell) \}\,.
\end{align}
From the proof in \cite{Gromov:2020fwh} we know that this determinant will expand into a sum over transfer matrix eigenvalues $\tau_{[\mu^\alpha_1,\mu^\alpha_2]}(\theta_\alpha)$. Together with the localization property and similar logic to the previous section, we can deduce that the determinant in \eqref{eqn:loc3} corresponds to an overlap $\langle \mathcal{K}_\kappa^\ell \otimes K^{L-\ell}|\Psi\rangle$. More precisely, we have
\begin{align}
   & \langle \mathcal{K}_\kappa^\ell \otimes K^{L-\ell}|\Psi\rangle = \langle K^L|\Psi\rangle \\ \nn
   & \mspace{20mu} \times\Theta_\ell \prod_{\gamma=1}^\ell \frac{\kappa^{1+2i\theta_\gamma}}{Q_{12}^-(\theta_\gamma)} \det_{(a,\alpha),(b,\beta)} \dab{\kappa^{-iu} u^{\beta-1}Q_a^{[b]}}_{\ell,\alpha}\,.
\end{align}
The precise form of the vector $\langle\mathcal{K}_\kappa|$ will depend on the choice of $K$. We make the choice
$\langle K| =\langle \mathcal{W}_\omega|$ and hence for this choice define $\langle \mathcal{W}_{\omega,\kappa}|:=c_\kappa \langle \mathcal{K}_{\kappa}|$ where $c_\kappa = (1-\omega z_1 z_2)(1-\kappa z_1)(1-\kappa z_2)$ is a convenient normalization. After this we can write
\begin{align}
   & \langle \mathcal{K}_\kappa^\ell \otimes \mathcal{W}_\omega^{L-\ell}|\Psi\rangle \\ \nn
   \mspace{10mu} =&\Theta_{L,\ell}\prod_{\gamma=1}^L \frac{\omega^{1/2+i\theta_\gamma}}{Q^-_{12}(\theta_\gamma)} \det_{1\leq \alpha,\beta\leq L} \dab{u^{\beta-1} \omega^{-iu} Q_{12}}_{L,\alpha} \\ \nn
   & \mspace{10mu} \times \prod_{\gamma=1}^\ell\frac{\kappa^{1+2i\theta_\gamma}}{Q_{12}^-(\theta_\gamma)} \det_{(a,\alpha),(b,\beta)} \dab{\kappa^{-iu} u^{\beta-1}Q_a^{[b]}}_{\ell,\alpha}\,.
\end{align}
We can then write the overlap as a sum over SoV wave functions and deduce $\langle \mathcal{K}_\kappa|$ by considering $\ell=L=1$. We find
\begin{equation}
    \langle \mathcal{W}_{\omega,\kappa}|=\langle Z| e^{C_{13}\mathbb{E}_{13}}e^{C_{14}\mathbb{E}_{14}}e^{C_{23}\mathbb{E}_{23}}e^{C_{24}\mathbb{E}_{24}}
\end{equation}
where
\begin{equation}
    C_{ij} =\frac{1-\kappa z_j}{1-\kappa z_i}R_{ij}
\end{equation}
which is related to $p^{\omega,\kappa}$ of \eqref{eq:WAdef} by
\begin{equation}\label{eqn:pol2}
\begin{split}
       & C_{13} = \nu_2 p^{\omega,\kappa}_{13},\quad C_{23}=\nu_1 p^{\omega,\kappa}_{23},\\
   & C_{14}=\nu_1 p^{\omega,\kappa}_{14},\quad C_{24}=\nu_2 p^{\omega,\kappa}_{24}\,.
\end{split}
\end{equation}

In the homogeneous limit $\Theta_{L,\ell}\rightarrow 1$ and $\dab{\cdot}\rightarrow \langle \cdot\rangle$ and hence
\begin{equation}
\begin{split}
   & \langle \mathcal{W}_{\omega,\kappa}^\ell \otimes \mathcal{W}_\omega^{L-\ell}|\Psi\rangle \nn \\
   & \mspace{20mu} =\frac{1}{Q_{12}(-\tfrac{i}{2})^{L+\ell}} \det_{1\leq \alpha,\beta\leq L} \langle u^{\beta-1} \omega^{-iu} Q_{12}\rangle_{L,\alpha}\\
   & \mspace{20mu} \times\det_{(a,\alpha),(b,\beta)} \langle\kappa^{-iu} u^{\beta-1}Q_a^{[b]}\rangle_{\ell,\alpha} \\
   & \times \omega^{L/2}\kappa^\ell (1-\omega z_1 z_2)^{-L} (1-\kappa z_1)^{-\ell}(1-\kappa z_2)^{-\ell}\,.
   \label{eq:overlapHomogeneous}
\end{split}
\end{equation}
As a check of the formula, one can take $|\Psi\rangle = |Z^L\rangle$ and accordingly $Q_{12}(u) = z_1^{-i u}z_2^{-i u}$. The right-hand-side then evaluates to~$1$, as expected.

To simplify the expression a bit, we use the notation from the main text and define $\mathcal{Q}_j = Q_j / \sqrt{Q_{12}(-i/2)}$, $\mathcal{Q}_{12} = Q_{12}/Q_{12}(-i/2)$, and
\begin{equation}
     \mathbb{W}_\omega=
\det_{1\leq\alpha,\beta\leq L}
    \langle \omega^{-iu}u^{\beta-1} \mathcal{Q}_{12} \rangle_{L,\alpha}
    \,,
\end{equation}
and
\begin{equation}
    \mathbb{A}_{\ell,\kappa} =
   \det_{(a,\alpha),(b,\beta)} \langle\kappa^{-iu} u^{\beta-1}\mathcal{Q}_a^{[b]}\rangle_{\ell,\alpha}\,.
\end{equation}
Then we have
\begin{equation}\label{eqn:finaloverlap}
\begin{split}
   & \langle \mathcal{W}_{\omega,\kappa}^\ell \otimes \mathcal{W}_\omega^{L-\ell}|\Psi\rangle = \mathbb{W}_\omega\times \mathbb{A}_{\ell}\\
   & \times \omega^{L/2}\kappa^\ell (1-\omega z_1 z_2)^{-L} (1-\kappa z_1)^{-\ell}(1-\kappa z_2)^{-\ell}\,.
\end{split}
\end{equation}

\section{The Complete Structure Constant}
\label{app:FullResult}

We now have all of the ingredients necessary to compute the twisted structure constant
\begin{equation}
    C_\ell = \frac{\langle \mathcal{W}_{\omega,\kappa}^\ell \mathcal{W}_\omega^{L-\ell}|\Psi\rangle}{||\Psi||}\,.
\end{equation}
By combining the result \eqref{eqn:finaloverlap} for the numerator with the expression \eqref{eqn:finalnorm} for the norm in the denominator we finally obtain
\begin{equation}
    C_\ell =\frac{\mathcal{N}_\ell \times \mathbb{W}_\omega  \times \mathbb{A}_{\ell,\kappa}}{(\mathcal{N}\, \mathbb{B})^{1/2}}
\end{equation}
where
\begin{equation}
    \mathcal{N} = (z_{13}z_{23}z_{14}z_{24})^{-L}\left(-\frac{\nu_1}{\nu_1^*}\right)^{K+M+\bar{K}} \left(\frac{\nu_2}{\nu_2^*}\right)^{K+\bar{K}}
\end{equation}
and
\begin{equation}
    \mathcal{N}_\ell = \frac{\omega^{L/2}\kappa^\ell (1-\omega z_1 z_2)^{-L} (1-\kappa z_1)^{-\ell}(1-\kappa z_2)^{-\ell}}{|p_{13}|^{2(K-\bar{K})}|p_{14}|^{2\bar{K}}|p_{23}|^{2(M-K)}}\,.
\end{equation}
Taking the absolute value we obtain \eqref{CSoV}.

\section{From hexagons to SoV}
\label{AppHexagons}

In this section we show how the hexagon sum over partitions can be written in an alternative determinant representation. The derivation proceeds in three steps: first, the partition sum is rewritten in an operator form; secondly,
Cauchy determinant identities and Bethe equations are used to exchange shifts in Bethe roots
for shifts acting on inhomogeneities; finally, we rewrite the resulting expression as a determinant, reproducing

\paragraph{Hexagon formulation.}

The structure constant is
\begin{align}
    C_{\ell}^2  &= \frac{\langle \textbf{v}|\textbf{w}\rangle^2 }{\langle \textbf{u}|\textbf{u}\rangle }\mathcal{A}_\ell^2
    \label{CHexagonsApp}\,,
\end{align}
where $\langle \textbf{u}|\textbf{u}\rangle$ and $\langle \textbf{v}|\textbf{w}\rangle$ are respectively the Gaudin norm and the wing norm, defined below
\begin{align}
    \langle \textbf{u}|\textbf{u}\rangle &= \prod_{i \neq j}^{M}\frac{1}{h(u_i,u_j)} \prod_{i \neq j}^{K} \frac{1}{h(v_i,v_j)}\prod_{i \neq j}^{\bar{K}}\frac{1}{h(w_i,w_j)} \times \nonumber \\
    &\times\begin{vmatrix}
        \partial_{u_i} \phi(u_j) & \partial_{u_i} \phi(v_j) & \partial_{u_i} \phi(w_j) \\
        \partial_{v_i} \phi(u_j) & \partial_{v_i} \phi(v_j) & \partial_{v_i} \phi(w_j) \\
        \partial_{w_i} \phi(u_j) & \partial_{w_i} \phi(v_j) & \partial_{w_i} \phi(w_j)
    \end{vmatrix}\,,
\end{align}
\begin{equation}
    \langle \textbf{v}|\textbf{w}\rangle  =  \prod_{i \neq j}^{K}\frac{1}{h(v_i,v_j)} \det \partial_{v_i} \phi(v_j)
\end{equation}
where
\begin{equation}
    h(u,v) = \frac{u-v}{u-v+i},
\end{equation}
and the phases $\phi$ are defined by the Bethe equations, which can be written in the form
\begin{align}
    e^{i\phi(u_j)} & \equiv \frac{z_2}{z_3} \left(\frac{u_j-\frac{i}{2}}{u_j+\frac{i}{2}}\right)^{L}\prod_{k \neq j}^{M}\frac{u_j-u_k+i}{u_j-u_k-i} \times \\
    &  \prod_{k=1}^{K}\frac{u_j-v_k-\frac{i}{2}}{u_j-v_k+\frac{i}{2}}\prod_{k=1}^{\bar{K}}\frac{u_j-w_k-\frac{i}{2}}{u_j-w_k+\frac{i}{2}} = 1 \\
    e^{i\phi(v_j)} & \equiv \frac{z_1}{z_2} \prod_{k \neq j}^{K}\frac{v_j-v_k+i}{v_j-v_k-i}\prod_{k=1}^{M}\frac{v_j-u_k-\frac{i}{2}}{v_j-u_k+\frac{i}{2}} = 1 \\
    e^{i\phi(w_j)} & \equiv \frac{z_3}{z_4}\prod_{k \neq j}^{\bar{K}}\frac{w_j-w_k+i}{w_j-w_k-i}\prod_{k=1}^{M}\frac{w_j-u_k-\frac{i}{2}}{w_j-u_k+\frac{i}{2}} = 1 \,,
\end{align}
Finally, $\mathcal{A}_\ell$ is the hexagon form factor we detail below.

\paragraph{Sum over partitions.}

After introducing a twist parameter $\tau$ and inhomogeneities $\theta_j$,
the hexagon expression takes the form
\begin{equation}
\mathcal{A}_{\ell} =
\sum_{\alpha \cup \bar{\alpha} = \mathbf{u}}
(-\tau)^{|\alpha|}
\prod_{j \in \alpha}f(u_j)a_\ell(u_j)
\prod_{i \in \alpha, j \in \bar{\alpha}}\frac{1}{h(u_i,u_j)} ,
\end{equation}
where
\begin{align}
&f(u) = \prod_{k=1}^K
\frac{u-v_k-i/2}{u-v_k+i/2},
\\
&a_\ell(u) = \prod_{j=1}^{\ell}
\frac{u-\theta_j-i/2}{u-\theta_j+i/2}.
\end{align}

The original quantity $\mathcal{A}$ is recovered in the homogeneous
($\theta_j\rightarrow0$) and untwisted ($\tau\rightarrow1$) limit.

\medskip

It is convenient to rewrite the sum over partitions in an operator form.
Introducing the Vandermonde determinant
\[
\Delta_{\mathbf{u}}=\prod_{i<j}(u_i-u_j),
\]
and Baxter polynomials
\[
Q_X(u)=\prod_{x\in X}(u-x),
\]
the expression becomes, as noted in \cite{Kostov:2012yq},
\begin{align}
\mathcal{A} =
\frac{Q^-_\theta(\mathbf{u})Q^-_{\mathbf{v}}(\mathbf{u})}{\Delta_\mathbf{u}}
\prod_{m=1}^M\left(1-\tau\mathbb{D}^{+1}_{u_m}\right)
\frac{\Delta_\mathbf{u}}{Q^-_\theta(\mathbf{u})Q^-_{\mathbf{v}}(\mathbf{u})}.
\end{align}
Now we'd like to exchange the shifts in $u$'s by shifts in $\theta$'s. We start by multiplying and dividing the expression by $\Delta(\mathbf{v} \cup \theta)$ . One obtains
\begin{align}
\mathcal{A} =
\frac{Q^-_\theta(\mathbf{u})Q^-_{\mathbf{v}}(\mathbf{u})}
{\Delta(\mathbf{u})\Delta(\mathbf{v} \cup \theta)}
\prod_{m=1}^M\left(1-\tau\mathbb{D}^{+1}_{u_m}\right)
\frac{\Delta(\mathbf{u})\Delta(\mathbf{v} \cup \theta)}
{Q^-_\theta(\mathbf{u})Q^-_{\mathbf{v}}(\mathbf{u})}.
\end{align}
with the shift operator $\mathbb{D}^{\pm 1}_x f(x) = f(x\pm i)$. When the number of Bethe roots satisfies $M = \ell + K$, the factor after (or before)
the parenthesis coincides with a Cauchy determinant,
\[
\det_{jk}\frac{1}{x_j-y_k-i/2},
\]
with $\mathbf{x}=\mathbf{u}$ and $\mathbf{y}=\mathbf{v} \cup \theta$.
This representation allows shifts acting on the variables $u_j$
to be exchanged for shifts acting on $v_k$ and $\theta_j$.

If $M \neq \ell + K$, the mismatch can be compensated by formally adding
Bethe roots at infinity. This produces an overall factor
$(\tau-1)^{M-\ell-K}$ and leads to
\begin{multline}
\mathcal{A} = (1-\tau)^{M-K-\ell}
\nonumber\times \\
\frac{Q^-_{\theta\cup \mathbf{v}}(\mathbf{u})}
{\Delta_{\mathbf{v} \cup \theta}}
\prod_{j=1}^\ell\left(1-\tau\mathbb{D}^{-1}_{\theta_j}\right)
\prod_{k=1}^K\left(1-\tau\mathbb{D}^{-1}_{v_k}\right)
\frac{\Delta_{\mathbf{v} \cup \theta}}
{Q^-_{\theta\cup \mathbf{v}}(\mathbf{u})}
\end{multline}

At this stage it is convenient to use the Bethe equations satisfied by
the auxiliary roots,
\begin{equation}
\frac{Q^+_{\mathbf{u}}(v_j)}{Q^-_{\mathbf{u}}(v_j)} =
\frac{z_1}{z_2}\prod_{k \neq j}^{K}
\frac{v_j-v_k+i}{v_j-v_k-i}.
\end{equation}
together with the same Cauchy determinant trick and the actual twisted Q-functions $Q_{12} = (z_1 z_2)^{-i u} Q_\mathbf{u}$ and $Q_{1} = z_1^{-i u} Q_\mathbf{v}$ to write
\begin{multline}
\mathcal{A} = \left(1-\kappa z_2\right)^{\lambda_2-\ell}
\left(1-\kappa z_1\right)^{\lambda_1-\ell}\times \\
\frac{1}{\Delta_\theta}
\prod_{j=1}^\ell
\left(
1-
\kappa g_2(\theta_j)
\mathbb{D}^{+1}_{\theta_j}
\right)
\left(
1-
\kappa g_1(\theta_j)
\mathbb{D}^{+1}_{\theta_j}
\right)
\Delta_\theta,
\end{multline}
with $\kappa = \tau/z_2$ and the $f$'s are almost the quantum eigenvalues $\Lambda$ \eqref{eqn:quantumeigenvalues} evaluated at $\theta_j$:
\begin{align}
    g_1 = \frac{Q_1^{++}}{Q_1} = \frac{\Lambda_1}{Q_\theta}\quad \text{and} \quad g_2 = \frac{Q_{12}^{+}}{Q_{12}^-}\frac{Q_1^{--}}{Q_1}= \frac{\Lambda_2}{Q_\theta}\,.
\end{align}

\paragraph{SoV wave functions and determinants.}

Notice we can also expand the product of the two parenthesis to obtain
\begin{multline}
\mathcal{A} = \left(1-\kappa z_2\right)^{\lambda_2-\ell}
\left(1-\kappa z_1\right)^{\lambda_1
-\ell}
\frac{1}{\Delta_\theta}
\times \\
\prod_{j=1}^\ell
\left(
 \tau_{[0,0]}(\theta_j)-\kappa \tau_{[1,0]}(\theta_j)
\mathbb{D}^{-1}_{\theta_j}+\kappa^2
\tau_{[1,1]}(\theta_j)
\mathbb{D}^{-2}_{\theta_j}
\right)
\Delta_\theta,
\label{eq:mathcalAfromTs}
\end{multline}
Interestingly, the parenthesis is equal to the determinant
\begin{align}
    \frac{1}{Q_{12}^-}\det\begin{pmatrix}
         Q_1-\kappa Q_1^{++} \mathbb{D}^{-1}_\theta&Q_1^{--}-\kappa Q_1\mathbb{D}^{-1}_\theta   \\
        Q_2-\kappa Q_2^{++}\mathbb{D}^{-1}_\theta&Q_2^{--}-\kappa Q_2\mathbb{D}^{-1}_\theta
    \end{pmatrix}
\end{align}
That fact, together with the determinant representation of the Vandermonde $\Delta_\theta=\det_{\alpha,\beta} \theta_\alpha^{\beta-1}$ imply the second line of \eqref{eq:mathcalAfromTs} can be written as a $2\ell \times 2\ell$ determinant made from $2\times2$ blocks
\begin{equation}
    \mathcal{A}_\ell = \frac{1}{\Delta_\theta^2\prod_{\alpha=1}^L Q^-_{12}(\theta_\alpha)}\det_{1\leq\alpha,\beta\leq L} \begin{pmatrix}
        [Q_1^+]_{\alpha,\beta} & [Q_1^-]_{\alpha,\beta} \\
        [Q_2^+]_{\alpha,\beta} & [Q_2^-]_{\alpha,\beta}
    \end{pmatrix}
\end{equation}
with
\begin{align}
    [f]_{\alpha,\beta}&= f^-(\theta_\alpha)\, \theta_\alpha
^{\beta-1}\  - \kappa f^+(\theta_\alpha) \, \left(\theta_\alpha-i\right)^{\beta-1}
\end{align}
By elementary row and column operations, we can go from this determinant to the one in \eqref{eqn:Adet1} once the integrals there have been computed by residues. That step completes the identification between \eqref{MapmathcalA} of $\mathbb{A}_{\ell,\kappa}$, defined by SoV, and $\mathcal{A}_\ell$ defined via Hexagons.

\section{Reduction to LR symmetric states and \texorpdfstring{$\grp{SU}(2)$}{SU(2)} states}

In this section we show how the Gaudin norm and DVD amplitudes in equations \eqref{BuSoV} and \eqref{AlSoV} simplify for operators in the $\grp{SU}(2)$ sector.

The Gaudin norm is related to the scalar product of two Bethe states by the following formula
\begin{align}
    \mathbb{B} = \frac{\prod_{j<k} z_{j k}^{\lambda_{j k}}}{\left(Q_{12}\left(-\frac{i}{2}\right) Q^{34}\left(\frac{i}{2}\right) \right)^{2L}} \det \langle Q_{j}^{[a]} u^{\beta-1} Q_{k}^{[b]} \rangle_{L,\alpha}
    \label{eq:su4Gaudin}
\end{align}
with $\lambda_{j k} = \lambda_j - \lambda_k$.

\paragraph{Gaudin norm for LR symmetric states.}

For LR-symmetric states, we have $Q^3, Q^4 = Q_2,Q_1$, respectively. For such states the determinant in \eqref{eq:su4Gaudin} factorizes into two pieces:
\begin{multline}
    \mathbb{B}\big|_{\text{LR}}= 2^L\frac{\prod_{j<k} z_{j k}^{\lambda_{j k}}}{\left(Q_{12}\left(\frac{i}{2}\right) Q_{12}\left(-\frac{i}{2}\right)  \right)^{2L}}\times \det_{1\leq\alpha,\beta\leq L}
        \langle Q_{12}\rangle_{\alpha,\beta} \\ \times \det_{1\leq\alpha,\beta\leq L} \begin{pmatrix}
        \langle Q^+_1 Q_2^+ \rangle_{\alpha,\beta} & \langle \{Q_1,Q_2\} \rangle_{\alpha,\beta} & \langle Q_1^- Q_2^-\rangle_{\alpha,\beta} \\
         \langle Q^+_1 Q_1^+ \rangle_{\alpha,\beta} & \langle Q^+_1 Q_1^- \rangle_{\alpha,\beta} & \langle Q^-_1 Q_1^- \rangle_{\alpha,\beta} \\  \langle Q^+_2 Q_2^+ \rangle_{\alpha,\beta} & \langle Q^+_2 Q_2^- \rangle_{\alpha,\beta} & \langle Q^-_2 Q_2^- \rangle_{\alpha,\beta}
    \end{pmatrix}
\end{multline}
where $\{P,Q\} = \frac{1}{2} \left(P^+ Q^- + P^- Q^+\right)$.

\paragraph{Gaudin norm for \texorpdfstring{$\alg{su}(2)$}{su(2)} states.}

For $\grp{SU}(2)$, we have $Q_1 = z_1^{i u}$. Therefore, $Q_1^+ = \frac{1}{z_1} Q_1^-$. Also remember $Q_{12} = Q_1^+ Q_2^- - Q_1^- Q_2^+$.
Then, we can rewrite the second determinant above as
\begin{align}
    \det_{1\leq\alpha,\beta\leq L} \begin{pmatrix}
        \langle Q^-_1 Q_2^+ \rangle_{\alpha,\beta} & \langle \{Q_1,Q_2\} \rangle_{\alpha,\beta} & \langle Q_1^+ Q_2^-\rangle_{\alpha,\beta} \\
         \langle Q^-_1 Q_1^+ \rangle_{\alpha,\beta} & \langle Q^+_1 Q_1^- \rangle_{\alpha,\beta} & \langle Q^+_1 Q_1^- \rangle_{\alpha,\beta} \\  z_1\langle Q^+_2 Q_2^+ \rangle_{\alpha,\beta} & \langle Q^+_2 Q_2^- \rangle_{\alpha,\beta} & \frac{1}{z_1}\langle Q^-_2 Q_2^- \rangle_{\alpha,\beta}
    \end{pmatrix}
\end{align}
    By subtracting half the first and third columns from the second column, we get:
\begin{multline}
    \left(\frac{1}{2}\right)^L\det_{1\leq\alpha,\beta\leq L}  \langle z_1^{-2iu}\,Q_{12} Q_{12} \, \rangle_{\alpha,\beta} \Big|_{\grp{SU(2)}}\\ \times
    \det_{1\leq\alpha,\beta\leq L}  \langle \,Q_{12}  \, \rangle_{\alpha,\beta} \times
    \det_{1\leq\alpha,\beta\leq L}  \langle \,z_1^{2 i u}  \, \rangle_{\alpha,\beta}
\end{multline}
Plugging that back in the original expression, one can verify:
\begin{multline}
    \mathbb{B}\big|_{\grp{SU(2)}} = \frac{\left(z_{12}^{-M} z_{13}^{-(L-M)} \det_{1\leq\alpha,\beta\leq L}
        \langle Q_{12}\rangle_{\alpha,\beta}\right)^2 }{\left(Q_{12}\left(\frac{i}{2}\right) Q_{12}\left(-\frac{i}{2}\right)  \right)^{L}}\\\times  \frac{ z_{23}^{-(L-2M)}\det_{1\leq\alpha,\beta\leq L}  \langle z_1^{2iu}\,Q_{12} Q_{12} \, \rangle_{\alpha,\beta}}{\left(Q_{12}\left(\frac{i}{2}\right) Q_{12}\left(-\frac{i}{2}\right)  \right)^{L}}
\end{multline}
The first factor is equal to the wing Gaudin, which for $\alg{su}(2)$ states is equal to $1$. Setting $z_1=1$ in the second factor, we get:
\begin{align}
    \mathbb{B}{\big|_{\grp{SU(2)}}} = \frac{\left(z_2-\frac{1}{z_2}\right)^{2M-L}}{\left(Q_{12}\left(\frac{i}{2}\right) Q_{12}\left(-\frac{i}{2}\right)  \right)^{L}}\times\det_{1\leq\alpha,\beta\leq L}  \langle Q_{12} Q_{12} \rangle_{\alpha,\beta}
\end{align}

\paragraph{DVD amplitude for \texorpdfstring{$\grp{SU}(2)$}{SU(2)} states.}

We can do the same for the DVD amplitude. We start with
\begin{multline}
    \mathcal{A}_\ell = \left(1-\kappa z_2\right)^{\lambda_2-\ell}
\left(1-\kappa z_1\right)^{\lambda_1-\ell}\times \\\det_{1\leq\alpha,\beta\leq L} \begin{pmatrix}
        \langle \kappa^{-i u} {Q_1}^+ \rangle_{\alpha,\beta} & \langle \kappa^{-i u} {Q_1}^- \rangle_{\alpha,\beta} \\  \langle \kappa^{-i u} {Q_2}^+ \rangle_{\alpha,\beta} & \langle \kappa^{-i u} {Q_2}^- \rangle_{\alpha,\beta}
    \end{pmatrix}
\end{multline}
Again, in the $\alg{su}(2)$ sector we have $Q_1 = z_1^{-iu}$ and $\lambda_1 = 0$. Using that and combining columns, we get:
\begin{multline}
    \mathcal{A}_\ell\big|_{\grp{SU(2)}} = \left(1-\kappa z_2\right)^{M-\ell}
\left(1-\kappa z_1\right)^{-\ell}\times \\\det_{1\leq\alpha,\beta\leq L} \begin{pmatrix}
        \langle (\kappa z_1)^{-i u}  \rangle_{\alpha,\beta} & 0 \\  \langle \kappa^{-i u} z_1^{-\frac{1}{2}}{Q_2}^+ \rangle_{\alpha,\beta} & \langle \kappa^{-i u} z_1^{i u} Q_{12} \rangle_{\alpha,\beta}
    \end{pmatrix}
\end{multline}
Setting $z_1 = 1$ and using $\det_{1\leq\alpha,\beta\leq L} \langle (\kappa z_1)^{-i u}  \rangle_{\alpha,\beta} = (1-\kappa z_1)^\ell$
\begin{align}
    \mathcal{A}_\ell \big|_{\grp{SU(2)}}= \left(1-\kappa z_2\right)^{M-\ell}\times \det_{1\leq\alpha,\beta\leq L}\langle \kappa^{-i u} Q_{12} \rangle_{\alpha,\beta}
\end{align}

\section{Proof of Localization Property}
\label{sec:localization}

Now we will prove the localization property \eqref{eqn:localization}. Suppose we have a transfer matrix $t_+(u)$ of length $L$ built with a twist matrix $g={\rm diag}(z_1,z_2,z_3,z_4)$, so that
\begin{equation}
    t_+(u) = \mathcal{L}_{i_1 i_2}(u-\theta_1)\dots\mathcal{L}_{i_L i_{L+1}}(u-\theta_L)g_{i_{L+1}i_1}
\end{equation}
where all indices $i_\alpha$ are summed over $\set{1,2,3,4}$. Under global $\grp{GL}(4)$ rotations $K_{\textbf{6}}$ in the physical space, the transfer matrix transforms in a very simple way
\begin{equation}
    K_{\textbf{6}} t_+(u) K^{-1}_{\textbf{6}} = t^K_+(u)
\end{equation}
where $t^K_+(u)$ is obtained from $t_+(u)$ by replacing $g\rightarrow
K_{\textbf{4}} g K^{-1}_{\textbf{4}}$.%
\footnote{This follows from the symmetry of the Lax operator~$\mathcal{L}$: For
$K\in\grp{GL}_4$, let $K_{\mathbold{6}}$ be its representation on the
six-dimensional physical space, and $K_{\mathbold{4}}$ its
representation on the auxiliary space (indices $j,k$ on
$\mathcal{L}_{jk}$~\eqref{eq:LaxOp}). Then
$K_{\mathbold{6}}\mathcal{L}K_{\mathbold{6}}^{-1}=K_{\mathbold{4}}^{-1}\mathcal{L}K_{\mathbold{4}}$.}
Hence, to prove the property \eqref{eqn:localization} we just need to show
\begin{equation}\label{eqn:genlocalize}
    \langle Z^L|\prod_{\alpha=1}^\ell \frac{t_+^K(\theta_\alpha)}{Q_\theta^{[-2]}(\theta_\alpha)} =  \langle Z^L| \prod_{\alpha=1}^\ell \frac{t_+^{K(\ell)}(\theta_\alpha)}{Q_{\theta,\ell}^{[-2]}(\theta_\alpha)}\,.
\end{equation}
We will show the proof for $L=2$ and $\ell=1$ which is enough to see the general logic. For convenience we will denote $G=K_{\textbf{4}} g K^{-1}_{\textbf{4}}$.

We need the following property of the Lax operator
\begin{equation}
\begin{split}
    & \langle Z|\mathcal{L}_{jj}(u) =(u-i)\langle Z|,\quad j=1,2 \\
    & \langle Z|\mathcal{L}_{jj}(u) =u\langle Z|,\quad j=3,4
\end{split}
\end{equation}
together with
\begin{equation}
    \langle Z|\mathcal{L}_{jk}(u) = 0,\quad j<k\,,
\end{equation}
which are just the highest-weight conditions, as well as
\begin{equation}
    \langle Z| \mathcal{L}_{21}(u)=0,\quad \langle Z| \mathcal{L}_{43}(u)=0\,,
\end{equation}
which is a bonus property of the six-dimensional representation of $\grp{GL}(4)$. Now we act with $t_+^K(\theta_1)$ on $\langle Z^2| = \bra{Z} \otimes \bra{Z}$, producing
\begin{equation}
    \langle Z|\mathcal{L}_{ij}(0)\otimes \langle Z|\mathcal{L}_{jk}(\theta_1-\theta_2)G_{ki}
\end{equation}
with all indices summed over. Applying the above properties of the Lax operator acting on $\langle Z|$, it is easy to check that this sum reduces to
\begin{align}
    \bra{Z^2}t_+^K(\theta_1)
    &= (\theta_1-\theta_2-i) \langle Z| \mathcal{L}_{ij}(0)G_{ji}\otimes \langle Z| \nn \\
    &= Q_{\theta}^{[-2]}(\theta_1)\langle Z^2|\frac{t_+^{K(1)}(\theta_1)}{Q_{\theta,1}^{[-2]}(\theta_1)}\,,
\end{align}
immediately yielding \eqref{eqn:genlocalize}. An identical procedure can be carried out for arbitrary length $L$, as well as the fused transfer matrices $t_0(u)$ and $t_-(u)$, thanks to the $\alg{gl}(4)$ covariance of the fusion procedure. This completes the derivation.

\section{Untwisting \texorpdfstring{$\grp{SU}(2)$}{SU(2)}}
\label{AppUntwist}

In this section we give more details on how the untwisted inner product can be obtained from the twisted inner product in the $\grp{SU}(2)$ sector.

Our starting point is the inner product
\begin{equation}
    B = \det_{1\leq \alpha,\beta \leq L}\langle Q_{12}(u) u^{\beta-1} Q_{12}(u)  \rangle_{L,\alpha}
    \label{su2ToUntwist}
\end{equation}
A typical matrix element of the determinant above is
\begin{align}
    \langle Q_{12}^2 u^{\beta-1} \rangle_{L,\alpha} = \oint \frac{du}{(-2\pi)^{\alpha}}\frac{e^{2\phi u}P_{M}(u)e^{2\pi u(\alpha-1)}}{\left(u-\frac{i}{2}\right)^{L}\left(u+\frac{i}{2}\right)^{L}}
    \label{matrixElementTwist}
\end{align}
where $ z_1 z_2 \equiv e^{i\phi}$ and $P_M$ is a polynomial of degree $M = 2(\lambda_1+\lambda_2) +\beta-1$, which accounts for the excitations of the two Q-functions as well as the powers of $u^{\beta-1}$ entering the matrix elements.

For highest-weight states, $P(u)$ is finite in the untwisting limit ($\phi\rightarrow0$), and the matrix element \eqref{matrixElementTwist} vanishes for $\alpha = 1$ while being generically non-zero for $\alpha > 1$. To see this vanishing behavior, one can close the contour at infinity and expand the integrand for large values of $u$:
\begin{align}
    \langle Q_{12}^2 u^{\beta-1} \rangle_{L,1} &\simeq -\oint_{\infty} \frac{du}{2\pi} \frac{e^{2\phi u}}{u^{2L-M}} = \nonumber \\
    &= -i\frac{(2\phi)^{2L-M-1}}{(2L-M-1)!} \left(1 + \mathcal{O}(\phi)\right)\,.
    \label{HowDiverges}
\end{align}

Thus, in the untwisting limit, each element of the first row of $\eqref{matrixElementTwist}$ is going to zero with a power of the twist as $\phi^{2L-M-1} = \phi^{2(L-\lambda_1-\lambda_2)-\beta}$, resulting in a vanishing determinant in the inner product \eqref{su2ToUntwist}. Since the other rows of this matrix are generically not zero, the leading term of its determinant is captured by its most vanishing minor with $\beta = L$, resulting in
\begin{align}
    B & \overset{\phi \to 0}{\longrightarrow} i(-1)^L \frac{(2\phi)^{L-M_A-M_B}}{(L-M_A-M_B)!} \times \nonumber \\
    &\times \det_{1\leq \alpha,\beta\leq L-1} \langle Q_A(u) u^{\beta-1} Q_B(u) \rangle_{L,\alpha+1}
\end{align}
When computing physical quantities the twist prefactors $z_{i,j}$ cancel the vanishing power of the twist, resulting in a finite untwisted physical inner product, that now has one extra row and one extra line.

\section{Examples}
\label{AppExamples}

\paragraph{Twisted theory.}

For any given state and twists, the Q-functions can be computed from the Baxter equation \eqref{niceBaxter}. As a concrete example, consider the state with $L=2$, Dynkin labels $[1,0,1]$ and whose Q-functions are given by
\begin{align}
    Q_1(u) & = z_1^{-i u} \nonumber \\
    Q_2(u) & = z_2^{-i u} \left(u + i\frac{(z_1+z_2^2)}{(z_1-z_2)(1+z_2)} \right)
\end{align}
By inserting these Q-functions into the SoV inner products \eqref{BuSoV}, \eqref{BvSoV} and \eqref{AlSoV}, evaluating the matrix elements \eqref{matrixelements} via residues for generic values of $\omega$ and $\kappa$ we obtain the fully twisted structure constant \eqref{CSoV}:
\begin{align}
    |C_{\ell=1}| &= \frac{(1+\kappa)(z_2-\omega z_1)(1-z_1z_2)}{\sqrt{2z_2}(z_1-z_2)(1+z_2)(1-\omega z_1 z_2)}\,, \\
    |C_{\ell=2}| &= \frac{(\kappa-z_2)(z_2-\omega z_1)(1-z_1z_2)}{\sqrt{2z_2}(z_1-z_2)z_2(1+z_2)(1-\omega z_1 z_2)}\,,
\end{align}
The main result \eqref{CSoV} allows us to compute structure constants for any finite value of twist. To demonstrate the applicability and power of our SoV formalism, we present explicit computations for both generic twist angles and the more physically relevant $\mathbb{Z}_2$ and $\mathbb{Z}_4$ orbifold points.

For the particular where we untwist the $\omega,\kappa = 1$ and set the twist factors to the values:
\begin{align*}
    \mathbb{Z}_2:& \hspace{0.4em}  z_1 = 1, \hspace{0.3em} z_2 = e^{-i\frac{\pi}{2}}, \hspace{0.3em} z_3 = e^{+i\frac{\pi}{2}}, \hspace{0.3em} z_4 = 1. \\
    \mathbb{Z}_4:&   \hspace{0.3em} z_1 = e^{-i\frac{\pi}{4}}, \hspace{0.3em} z_2 = e^{-i\frac{3\pi}{4}}, \hspace{0.3em} z_3 = e^{+i\frac{3\pi}{2}}, \hspace{0.3em} z_4 = e^{+i\frac{\pi}{4}},
\end{align*}
we recover the $\mathbb{Z}_2$ and $\mathbb{Z}_4$ orbifold points of $\mathcal{N}=4$ super Yang--Mills. And consequently, the structure constant above reduces to
\begin{align}
    \mathbb{Z}_2: |C_{\ell=1}|^2 = 1 \qquad |C_{\ell=2}|^2 = \frac{3}{2}+\sqrt{2} \\
    \mathbb{Z}_4: |C_{\ell=1}|^2 = \frac{1}{2} \qquad |C_{\ell=2}|^2 = 1+\frac{1}{\sqrt{2}}
\end{align}

\paragraph{Untwisted theory.}

We start by setting $\omega=1$ and $\kappa = z_1$, while parametrizing the remaining twist as
\begin{equation}
    z_1 = e^{-i\phi}, \hspace{0.6em} z_2 = e^{-i{\phi}/{2}}, \hspace{0.6em} z_3 = e^{+i{\phi}/{2}}, \hspace{0.6em} z_4 = e^{+i\phi},
    \label{ZNumerics}
\end{equation}
then we solve the Baxter equation \eqref{niceBaxter} at small $\phi$, obtaining Q-functions as a power series in the twist parameter.

Let's consider two $L=3$ states with Dynkin labels $[0,1,0]$, one primary and one descendant. The primary Q-functions expanded around the untwisting limit ($\phi \to 0$ in \eqref{ZNumerics}) are given by
\begin{align*}
    Q_1(u) & = e^{-\phi u}\left(u+ \tfrac{\sqrt{3}}{5}-\tfrac{36}{125}\phi+\tfrac{163\sqrt{3}}{6250}\phi^2+\dots \right) \\
    Q_2(u) & = e^{-\frac{\phi}{2}u}\left(u+\tfrac{\sqrt{3}}{5}-\tfrac{16}{125}\phi + \tfrac{103\sqrt{3}}{6250}\phi^2+\dots\right)
\end{align*}
while the descendant Q-functions are
\begin{align*}
    Q_1(u) & = e^{-\phi u}\left(u+ \tfrac{15+\sqrt{33}}{4\phi}-\tfrac{\phi}{8}- \tfrac{385+57\sqrt{33}}{168960}\phi^3+\dots\right) \\
    Q_2(u) & = e^{-\frac{\phi}{2}u}\left(u+ \tfrac{1-\sqrt{33}}{2\phi}-\tfrac{\phi}{24}+ \tfrac{143-153\sqrt{33}}{506880}\phi^3+\dots\right)
\end{align*}

At this stage one can see the major difference between primary and descendants: primaries have finite Q-functions in the untwist limit, while the Q-functions of descendants diverge as we untwist.

Nonetheless, our formalism treats them in the exact same way. Inserting these Q-functions into \eqref{CSoV} and extracting the leading order in $\phi$, we obtain finite expressions for the untwisted structure constants of the primary
\begin{equation}
    C^{2}_{\ell=1} = \frac{1}{15}\,, \quad C^{2}_{\ell=2} = \frac{1}{15}\,,
    \label{Cprimary}
\end{equation}
as well as of the descendant
\begin{equation}
    C^{2}_{\ell=1} = \frac{11}{4}-\frac{41}{4\sqrt{33}}\,, \quad C^{2}_{\ell=2} = \frac{3}{8}-\frac{15}{8\sqrt{33}}\,.
    \label{CDesc}
\end{equation}

\bibliography{ref}

\end{document}